# Progress in wire fabrication of iron-based superconductors

Yanwei Ma[*]

Key Laboratory of Applied Superconductivity, Institute of Electrical Engineering,
Chinese Academy of Sciences, Beijing 100190, China

**Abstract:**

Iron-based superconductors, with $T_c$ values up to 55 K, are of great interest for applications, due to their lower anisotropies and ultrahigh upper critical fields. In the past 4 years, great progress has been made in the fabrication of iron-based superconducting wires and tapes using the powder-in-tube (PIT) processing method, including main three types of 122, 11, and 1111 iron-based parent compounds. In this article, an overview of the current state of development of iron-based superconducting wires and tapes is presented. We focus on the fabrication techniques used for 122 pnictide wires and tapes, with an emphasis on their meeting the critical current requirements for making high-performance conductors, such as a combination of using Ag sheath, addition element and optimized heat treatment to realize high $J_c$, *ex situ* process employed to reduce non-superconducting phases and to obtain a high relative density, and a texture control to improve grain connectivity. Of particular interest is that so far transport $J_c$ values above $10^4$ A/cm$^2$ at 4.2 K and 10 T are obtained in 122 type tapes, suggesting that they are prospective candidates for high-field applications. Finally, a perspective and future development of PIT pnictide wires are also given.

[*] E-mail: ywma@mail.iee.ac.cn

**Contents**

**1. Introduction**

In February 2008, the Hosono group in the Tokyo Institute of Technology discovered superconductivity at 26 K in the oxypnictide LaFeAsO$_{1-x}$F$_x$, representing a new class of high-$T_c$ superconductors [1]. Soon after the discovery, the superconducting transition temperature $T_c$ was raised up to 55 K by replacing La with other rare-earth elements [2], which is now called the '1111' phase. Like the CuO-layered high-$T_c$ cuprates, iron pnictides are also layered materials, with the superconductivity occurring in FeAs sheets, with the RE-O layers acting as a charge reservoir. Subsequently, three other series of new compounds have been identified: 122 phase (AEFe$_2$As$_2$ with AE = alkali or alkali earth) [3–5], 111 phase (AFeAs with A = alkali metal) [6] and 11 phase (FeSe or FeTe) [7]. These four types of compounds have a common structure unit, a square lattice of Fe$^{2+}$ ions, but with different $T_c$ values, as shown in figure 1. Among them, the three phases most relevant for wire applications are 1111, 122, and 11 types with a $T_c$ of 55, 38 and 8 K, respectively.

Compared with the Cu-based cuprates, the iron-based superconductors have metallic parent compounds, and the attractive low anisotropy, in general smaller than any cuprate, for example, about ~2 near $T_c$ in 122-type, while YBa$_2$Cu$_3$O$_{7-x}$(YBCO), is about 7–20 near the optimal doping and strongly dependent on the doping level [8]. Different from the cuprates with d-wave symmetry, the pnictides are s-wave-like, which is in principle not so detrimental to current transport across grain boundaries. Table I summarizes the features of iron-based superconductors in comparison with those of cuprates and MgB$_2$, the intermetallic compound superconductor with the highest $T_c$ of 39 K [9]. All of these materials have layered structures. Clearly, three unique properties are evident for iron pnictides: robustness to impurity doping, very high upper critical field, and low crystallographic anisotropy in physical performances, implying that iron pnictides are favorable for application to superconducting wires.

From the viewpoint of practical applications, such as magnets and cables, the development of a pnictide wire processing technique is indispensable. Iron-based superconductors are mechanically hard and brittle and, therefore, difficult to be deformed plastically into wires and tapes. The most common approach for developing wires from such brittle superconductors is the powder-in-tube (PIT) process [10], in which powders are packed in a metal tube, mechanically processed into a wire or tape form, and finally heat treated for reaction, like the case of Bi-2223 and $MgB_2$ [11-12]. Actually, shortly after the discovery of oxypnictide superconductors, the first fabrication of pnictide wire was reported by the authors' group, through applying PIT process for LaFeAsOF [13], SmFeAsOF [14], and (Sr, K)$Fe_2As_2$ [15] superconductors.

In the light of these early results and of the rapidly growing literature dealing with the preparation of iron-based superconducting conductors with the aim of assessing their potential for application, the feasibility of fabricating powder-in-tube processed pnictide wires needs to be explored in detail. The aim of this paper is to review the preparation procedures and the most recent results and developments of iron pnictide wires and tapes made by the PIT technique, paying particular attention to 122 type conductors. Various attempts to realize high quality wires will be described, including a choice of Ag sheath, addition techniques, heat treatment optimization, *ex situ* process, and a new texturing method. Finally, a future prospect in the development of iron-based superconducting wires will be briefly presented.

## 2. Superconducting properties of iron-based superconductors

The superconducting transition temperature, $T_c$, upper critical field, $H_{c2}$, and critical current density, $J_c$, are the three key properties for applications. Actually, since the discovery of pnictides, superconducting properties have been widely investigated, based on single crystal and bulk samples. The solid state reaction and the high pressure synthesis method are mainly used for the preparation of polycrystalline samples [10], while the flux method is widely used for making single crystals [16]. As for the 1111 family, the single crystal is difficult to be synthesized because of the high melting temperature (~1200 °C) and volatility of F element. So far, only several groups have reported to grow it but with small size of crystals [17, 18]. On the other hand, many single-crystal results were reported in the 122 system since larger crystals can be easily grown.

Upper critical field is an intrinsic property of a type II superconducting material in which the resistive transition ($\rho$-$T$) width increases with increasing magnetic field. The $H_{c2}$(0 K) can be calculated using the one band Werthamer–Helfand–Hohenberg (WHH) formula [19], $H_{c2}$(0 K) = -0.693$T_c$[$dH_{c2}/dT$]$_{Tc}$, where [$dH_{c2}/dT$]$_{Tc}$ is obtained from the slope of the curve $H_{c2}$ versus $T$ at $T_c$. It is noted that $H_{c2}$ values extrapolated using the WHH model are not valid for the low temperature range [20]. Hunte et al. reported an upper critical field of 65 T in polycrystalline La-1111 with $T_c$= 26 K by the first magnetoresistance measurements up to 45 T [21]. By replacing La with smaller rare earths like Nd and Sm, $T_c$ and $H_{c2}$(0 K) both increase [22], e.g., $H_{c2}$ was estimated to be higher than 100 T in bulk SmO$_{0.85}$F$_{0.15}$FeAs with $T_c$ of 46 K [23] and NdFeAsO$_{0.82}$F$_{0.18}$ with $T_c$ of 51 K [24]. Jia et al. have first grown single crystals of NdFeAsO$_{0.82}$F$_{0.18}$ with $T_c$ =47 K at ambient pressures [17], and the upper critical fields determined based on the WHH extrapolations were even $H_{c2}^{//ab}(0K) \approx 304$ T and $H_{c2}^{//c}(0K) \approx 62 - 70$ T. Further studies on single crystals of NdFeAsO$_{0.7}$F$_{0.3}$ at very high magnetic field up to 60 T, by Jaroszynski et al., also showed the 100–200 T field scale of $H_{c2}$ because of high slopes $H_{c2}$~3–20 T/K at $T_c$ [22]. The anisotropy evaluated as $\gamma_H = H_{c2}^{//ab}/H_{c2}^{\perp ab}$ is also strongly temperature-dependent, varying from 4 to 8. However, a different situation is observed in the 122 family. (Ba, K)Fe$_2$As$_2$ single crystals exhibited nearly isotropic $H_{c2}$, measured under pulsed magnetic fields, with values of the order of 60 T at zero temperature and anisotropy going from 2, close to $T_c$, down to 1 at 5 K [20]. Co-doped BaFe$_2$As$_2$ single crystal showed an anisotropy of 1−3 and upper critical field values of $H_{c2}^{//ab} = 20$ T and $H_{c2}^{//c} = 10$ T at 20 K, with $dH_{c2}/dT$=5 T/K [25]. After that, more results have appeared that the estimated $H_{c2}$(0 K) of Ba-122 single crystals based on the low-field $H_{c2}$(T) experiments exceeded 100 T at low temperatures[26-28]. Fang et al. determined the $H_{c2}$(0 K) as 47 T for single crystals of the Fe$_{1.11}$Te$_{0.6}$Se$_{0.4}$ ($T_c$≈14 K) using pulsed magnetic fields of up to 60 T [29].

As shown in figure 2, the new iron-based superconductors have rather high values of $H_{c2}$, quite comparable to the cuprates. It is clear that both NdFeAsO$_{0.7}$F$_{0.3}$ and Ba$_{0.55}$K$_{0.45}$Fe$_2$As$_2$ could in principle provide fields up to 40–50 T at 20 K [30]. As the less anisotropic Ba$_{0.55}$K$_{0.45}$Fe$_2$As$_2$ is with $1 < \gamma(T) < 2$, the conductors based on Ba-122 are more promising for high field magnets operating at 20 K.

From the application point of view, large critical current densities $J_c(T, H)$ are also required, preferably for magnetic fields higher than 5 T. For the 1111 class, Zhigadlo et al. reported a high magnetic $J_c$ of $2\times10^6$ A/cm$^2$ at 5 K on a SmFeAsO$_{1-x}$F$_x$ single crystal. Subsequent transport measurements revealed a promising $J_c$ of more than $10^6$ A/cm$^2$ at 5 K, 14 T and nearly isotropic critical current densities along all crystal directions [18]. For the Ba$_{0.6}$K$_{0.4}$Fe$_2$As$_2$ single crystals, Yang et al. reported significant fishtail peak effects and large current carrying capability up to over $10^6$ A/cm$^2$ at 5 K in a field of 9 T [31]. Similar fishtail peak behavior with $J_c$ of $4\times10^5$ A/cm$^2$ at 4.2 K was reported in Ba(Fe,Co)$_2$As$_2$ crystals [32]. Subsequently, many groups reported that the $J_c$ in single crystals of the Co-doped Ba(Fe$_{1-x}$Co$_x$)$_2$As$_2$ was over $10^5$ A/cm$^2$ at 5 K for $H//ab$ and $H//c$, respectively [25, 33-34]. The $J_c$ decreases with increasing magnetic field up to 1 T and after that become nearly field independent which is related to relatively high pinning potential and weakly anisotropic property.[34] In addition, pnictide thin film can also carry high current densities $J_c\sim10^6$ A/cm$^2$ at 4.2 K, quite similar to those of single crystals [9].

All these results show that iron-based superconductors exhibited extremely high $H_{c2}$, very low anisotropy, and rather large $J_c$ values with independent of the field at low temperatures, indicating good possibilities for magnet applications at 20-30 K, where the niobium-based superconductors cannot play a role owing to their lower $T_c$s.

As we know, wires for bulk applications are always based on polycrystalline materials, however, early studies showed that a global $J_c$ in polycrystalline samples has appeared just in the orders of $10^3$ A/cm$^2$ at 4 K in self-field [35-37]. It was reported that strong granularity of these compounds restricted global $J_c$ values to very low ones. Such granular behavior has also limited the properties of pnictide wires [13-15]. Microstructural studies have emphasized cracks, low density, grain boundary-wetting FeAs phase, and phase inhomogeneities which cause local suppression of the order parameter at grain boundaries as the main mechanisms responsible for current blocking in polycrystalline materials, which will be discussed more in the latter sections.

**3. Preparation of iron-based superconducting wires and tapes by the PIT method**

**3.1 The powder-in-tube process**

Iron-based superconductors are a relatively tough and hard phases, and thus cannot be plastically deformed. The only way to obtain a filamentary configuration is to start with powders that are packed in metallic tubes. Therefore, the powder-in-tube (PIT) method is a very convenient way of producing the iron-based superconducting wires, which is currently studied in many laboratories. The PIT process is very attractive from the aspect of applications, taking advantage of the low material costs and the relatively simple deformation techniques. Actually, the PIT process is often used for making electrical conductors from brittle superconducting materials such as $Nb_3Sn$ [38], $MgB_2$ [12], and ceramic Bi-cuprate superconductors [11].

The basic PIT process is schematically illustrated in figure 3. Generally, the PIT pnictide wire is fabricated by packing stoichiometric amounts of 11, 122 or 1111 powder particles into a metal tube under Ar atmosphere and sealed to form a billet. The billet is then swaged and drawn to wire composites and finally given a heat treatment under Ar [10]. Furthermore, the PIT technique is usually classified into two different processes: *in situ* and *ex situ*. In the former case, a mixture of the starting materials is packed into a metal tube and the reaction is performed within the final wire or tape after deformation, whereas the latter employs a precursor of synthesized superconducting material before filling into the metal tube.

It is worth noting that we also modified the PIT technique to synthesize wire-shaped bulk samples [39], whose major merits are safety and convenience. Actually, this one-step PIT method is quite effective and fast for exploring new iron-based superconductors, for instance, we have successfully discovered superconductivity at 34.7, 15.2 and 11.8 K in new $Eu_{0.7}Na_{0.3}Fe_2As_2$, Co-doped SmFeAsO and Ir-doped LaFeAsO compounds, respectively [5, 40-41], demonstrating that the one-step PIT synthesis process is unique and versatile and hence can be tailored easily for other rare earth derivatives of pnictide superconductors.

The following sections provide an overview of a few key advances that enabled the development of PIT pnictide wires.

### 3.2 The first fabrication of iron-based superconducting wires

In the early stage of pnictide wire development, one of the main challenges encountered when fabricating wires or tapes by the PIT process is the hardness and brittleness of the compound pnictide. On the other hand, the use of proper metal cladding is another critical issue because of the strong chemical reactivity of the pnictide at the high annealing temperatures of 900-1200°C for tens of hours. Thus, the choice of the metallic sheath has been reduced to those elements or metals showing little or no reaction with pnictide at this temperature range. Earlier sheath materials such as Nb, Ta and Fe have been tried for making metal–clad pnictide wires [13-15].

Just two months after the discovery of oxypnictides, Our group were the first to report La-1111 and Sm-1111 superconducting wires by the *in situ* PIT technique [13,14]. We mixed La(Sm), As, $LaF_3(SmF_3)$, Fe, and $Fe_2O_3$ as starting materials. A Ta tube or an Fe tube with an inner Ti sheath was used to prevent the reaction between the tubes and the 1111 type compounds. After packing, the tubes were then rotary swaged and drawn to wires of about 2.0 mm in diameter, and finally given a recovery heat treatment at the temperature 1180 °C for 45 h. The final Ta-sheathed $SmFeAsO_{1-x}F_x$ wires are shown in figure 4.

Figure 5 shows the temperature dependence of resistivity for the samples $SmO_{1-x}F_xFeAs$ after peeling off the Ta sheath. It shows a $T_c$ as high as 52 K, and a residual resistivity ratio RRR of 2.8 for $x$=0.35 sample. Magnetization measurements revealed that the core material of the wires had a self-field $J_c$ of approximately 4000 A/cm$^2$ at 5 K and a weak magnetic field dependence of $J_c$, as shown in figure 6, indicating an encouraging first step toward fabrication of practical wires.

Although Sm-1111 wire has high $T_c$ up to 52 K, however, it is difficult to be synthesized because the sintering temperature is quite high (~1200°C). Thus, the 122 type pnictides seem a better candidate for making wires, due to low annealing temperature used (~850°C) and no oxygen involved. The first superconducting PIT wires of the 122 type compound $Sr_{0.6}K_{0.4}Fe_2As_2$ were fabricated by Qi et al. using the Nb as sheath material [15], just after the report of 1111 type wires. The $T_c$ of the Sr-122 wires is confirmed to be as high as 35.3 K. Figure 7 shows magnetic field dependence of $J_c$ at different temperatures for Sr-122 wires, estimated from the magnetization hysteresis loops. Although the magnetic $J_c$ was rather low, 3700 A/cm$^2$ at 5 K, Sr-122 wires exhibited a weak $J_c$-field dependence behavior even the temperature is very close to $T_c$. The upper critical field $H_{c2}$(0 K) value could exceed 140 T [15], surpassing those of $MgB_2$ and all the low temperature superconductors.

### 3.3 The use of silver as the sheath material

As mentioned above, the first Sm- (or La-)1111 and Sr-122 superconducting wires [13-15] had magnetic global $J_c$ values of ~4000 A/cm$^2$ at 5 K, some ten times higher than in random polycrystalline cuprates, which are typically 100 A/cm$^2$ due to large intrinsic weak-linked effects [35, 42]. Thus it may suggest that the weak-linked effect at pnictide grain boundaries is less serious than in the cuprates. Unfortunately, subsequent transport measurements revealed that no transport critical current was observed for either 1111- or 122 type wires.

For the PIT process, the wires are usually subject to a sintering process at high temperatures, therefore we have to use sheath materials that are not reactive with pnictide. However, early efforts suffered from the reaction layer problem. In the previous study, Nb, Ta and Fe have been employed as sheath materials for wire fabrication. From the transverse and longitudinal cross-sections of a sintered pnictide wire as shown in figure 4a and 4b, a reaction layer with a thickness 10~30 μm between the superconductor core and the tube was clearly observed, although the thickness of reaction layer in Sr-122 wires [15] is smaller than that in Sm-1111 ones due to the lower sintering temperature. In order to get more information about the interfacial reaction, Zhang et al. have systematically studied the effect of various sheath materials on the microstructure and superconducting properties of Sm-1111 wires [43]. By means of the elemental maps and EDX spectra measurements, indeed all the sheaths such as Nb, Ta and Fe/Ti showed a markedly higher reaction with the pnictide due to the diffusion of As into the sheath, the interfacial reaction layer ranges from 60 to 200 μm thick. Meanwhile, this reaction consumed some of the As within the filament, leading to an increase in porosity. These results demonstrated that such a thick reaction layer formed between the core and the sheath would act as a big barrier for obtaining the transport $J_c$ values in the wire samples. Therefore a key issue is to find a valuable alternative to the above sheaths showing little or no reaction with pnictide.

In 2009, Wang et al. fabricated $Sr_{0.6}K_{0.4}Fe_2As_2$ wires and tapes by packing the raw material inside a Ag tube and made the first success in measuring the transport critical current in the 122 type wires [44]. For mechanical reinforcement, composite sheath material Ag-Fe was used. The achievement of using silver as the sheath material was a key breakthrough and finally solved the reaction problem for pnictide wire to observe the transport $J_c$. The transverse cross-sections of a typical $Sr_{0.6}K_{0.4}Fe_2As_2$/Ag/Fe wire and tape taken after heat treatments (900°C for 35 h) were shown in figure 8a. Both Ag/Fe and Sr-122/Ag interfaces were quite clear, no reaction layer was observed between the silver sheath and the superconducting core (figure 8b), indicating silver is benign in proximity to the compound at high temperatures. EDX line-scan further confirmed no diffusion of As or Sr into the volumes of Ag [45], which benefits superconducting properties of the superconducting core. Very soon, the use of a silver sheath also led to the first successful fabrication of F-doped Sm-1111 wires with a transport $J_c$ of 1300 A/cm$^2$ at 4.2 K, 0 T [46], which will be described later.

Regarding the first Ag-sheathed 122 tape samples, the $T_c$s were estimated to be 34 K, which is the same as that of the bulk material. Thanks to the Ag sheath, all the Sr-122 wire and tape samples have shown the ability to transport superconducting current, typically around several Amperes [44]. Transport critical currents were measured by a standard four-probe resistive method using a criterion of 1 μV/cm. As shown in figure 9, the self-field transport $J_c$ reached 500 A/cm$^2$ at 4.2 K, but $J_c$ falls to 30 A/cm$^2$ as 0.5 T is applied. The field dependence of $J_c$ in an increasing as well as a decreasing field was also characterized (see the inset of figure 9), clearly a hysteretic phenomenon has been observed, similar to that of sintered YBCO [47-48], indicating a weak-linked behavior for the obtained tapes. These results do support the earlier speculation that the absence of significant transport currents in the previous PIT wires was caused by extrinsic blocking effects [13-15].

Although the measured $I_c$ value was small for superconducting wire available for applications, it was the first report of the transport critical current density for a 122 pnictide superconducting wire [44]. In addition, Fujioka et al. claimed that Cu could work as a good sheath material since it reacted only very weakly with polycrystalline Sm-1111 after annealing at 1000°C for 20 h [49], but they did not show any critical current data for their wires. In conclusion, the use of silver as the sheath material was critical to the development of the PIT process for iron based superconductors, as it seems the only material that is chemically compatible with pnictide. Since then, Ag sheath has been widely applied to fabricate iron-based superconducting wires and tapes by other groups, such as NIMS, Florid State University, Tokyo University, etc. [50-52]

**4. 122 type wires and tapes**

The 122 family compounds have much lower anisotropy than the 1111 pnictides and high $T_c$ values close to that of $MgB_2$ [53]. They also show nearly isotropic superconductivity [20] and very high intrinsic pinning potential, which is weakly field dependent [27]. In particular, the 122 pnictides are rather stable phase, and the single phase can easily be obtained using a conventional solid-state reaction method. These unique features make the 122 superconductors more favorable for applications than other pnictide superconductors. Therefore, a rather significant research effort was directed toward fabricating superconducting 122 wires.

As noted in the previous section, Qi et al. first reported the fabricating wires of 122 type pnictides, Sr-122 superconducting wires by the *in situ* PIT method [15]. Subsequently, Wang et al. succeeded in producing a superconducting Sr-122 wire with the first transport $J_c$ property, using silver as a sheath material [44]. However, the 122 wires showed a weak-linked behavior and a rather low self-field $J_c$ of 500 A/cm$^2$ at low temperatures, which is much smaller compared to those of the single crystals and films [31-32, 54-55]. The low $J_c$-$H$ properties imply either very weak pinning or an imperfectly connected superconducting state coming from either the secondary impurity phases, or residual cracks and porosity. When examining these Sr-122/Ag tapes, we found large pores and poor intergrain connections for the samples [44], thus introducing a strong limitation to the flow of currents. The fact also corroborated by the previous results, in which FeAs phase wets many grain boundaries, thus interrupting grain-to-grain supercurrent paths, which are further degraded by extensive cracking [56]. Hence, how to decrease the porosity and to improve the grain connection are a critical problem to be resolved for improving $J_c$ values of iron based superconducting wires.

**4.1 Ag or Pb addition to improve the grain connectivity**

Chemical addition usually plays an important role in enhancing superconducting properties by promoting the crystallization of the superconducting phase, catalyzing the intergranular coupling of the superconducting grains or introducing pinning centers. For example, the irreversibility field and $J_c$ can be largely increased by carbon addition in $MgB_2$ and a $J_c$ enhancement was observed in Ag-added YBCO [57-59]. In the development of the 122 pnictide wires, another crucial advance was the introduction of additive elements such as Ag or Pb to improve the intergranular coupling of the superconducting grains, hence the enhanced $J_c$ properties.

The effect of Ag addition on the polycrystalline Sr-122 superconductor was first investigated by Wang et al. [60], who prepared *in situ* bulk samples containing between 0 and 20 wt.% silver powders. Sr filings, Fe powder, As and K pieces, with a ratio Sr : K : Fe : As = 0.6 : 0.44 : 2 : 2.2, were thoroughly ground in Ar atmosphere for more than 10 hours using a ball milling method.

X-ray diffraction (XRD) results showed that the Ag added samples consisted of $Sr_{0.6}K_{0.4}Fe_2As_2$ as the major phase, with some Ag and a small amount of impurity phases, which were identified as FeAs and AgSrAs. Resistivity drops at 35 K and vanishes at about 33 K for all samples, clearly Ag addition hardly affected the critical transition temperature. Most notably, significant enhancement of critical current density $J_c$ has been observed for the Ag added bulk samples, as shown in figure 10. The $J_c$ of 20 % Ag added samples at 5 K in 0 T is about $2.5 \times 10^4$ A/cm$^2$ and remains above $1.5 \times 10^3$ A/cm$^2$ beyond 6.5 T, twice as high as for the pure sample. Additionally, excellent $J_c$ of about $1.0 \times 10^4$ A/cm$^2$ at 20 K was achieved in the 20% Ag added samples [60].

Figure 11 shows scanning electron micrographs of the polished surface of the pure and Ag-added samples. The pure sample revealed light gray Sr-122 grains and a large amount of dark gray impurity phases (perhaps glassy phases, due to the absence from the XRD pattern). Note that these prominent impurity phases were frequently observed in RE-1111 polycrystals too. [35, 56] By contrast, some white Ag particles can be seen in the Ag-added samples, as shown in figure 11(c, d). Most interestingly, the dark gray glassy phases (wetting phase or liquid phase) were much reduced after Ag addition, indicating that Ag has the beneficial effect on the grain connectivity. TEM study showed that Ag addition can effectively suppress the formation of glassy phase as well as amorphous layer, leading to better connections between grains thus higher $J_c$ [60]. It is noted that the addition of Ag rarely changes the values of the lattice constants, meaning that Ag hardly enters into the 122 lattice [45].

Subsequently, the positive effect of Ag addition was further proved by the remanent magnetization results of Otabe et al. [61] Figure 12 shows the derivative of the remanent magnetic moment as a function of increasing applied field for the Sr-122 bulk with and without the benefit of Ag addition. For the Ag added bulks, two peaks were clearly observed. The low-field peak is related to the global critical current density while the higher field peak is connected to the local critical current density. However, only one high field peak was present in pure samples, which is caused by locally circulating currents with current loop size less than the powder size. It is known that two kinds of loops, intra-grain current loops and inter-grain current loops, contribute to the magnetization of granular superconductors [35, 36]. Based on the inter and intragranular $J_c$ deduced from the lower and higher peaks of the $dm_R/dH_m$, they found that the intergranular $J_c$ of the order of $10^3$ A/cm$^2$ at 5 K is observed in Ag-added specimen while the intergranular $J_c$ is too small to be observed in pure sample, indicating the strongly improved grain coupling made by Ag addition. Furthermore, they also estimated the filling factor for 122 pnictide polycrystalline samples, which is very important for practical applications [61].

Indeed, through Ag addition, we have successfully observed larger transport critical currents in the 122 pnictide wires and tapes by the *in situ* PIT process. At 4.2 K, a high transport $J_c$ of 1200 A/cm$^2$ in zero field and 100 A/cm$^2$ at 10 T have been obtained in the Ag-added tapes as shown in figure 13, more than two times larger than that of the pure tapes. The $J_c$ enhancement upon Ag addition was mainly due to the elimination of pores and enhanced connectivity (see figure 14). Soon after, a positive influence of the Ag addition on $J_c$ of 122 wires was also reported by other groups [50-51]. This earlier work demonstrated the feasibility of fabricating iron-based superconducting wires with high transport $J_c$ through the PIT method.

At the same time, we studied the effects of Pb addition on microstructure and superconducting properties of 122 polycrystalline samples [62]. We found that an enhancement of $J_c$ in pnictide bulks and tapes was achieved by Pb addition, as shown in figure 15. As is evident from the figure, magnetic $J_c$ in the entire field region was increased by Pb addition. A substantial improvement was obtained by increasing the Pb content up to 10%, while upon further increasing the Pb content, the $J_c$ decreased, because of too much non-superconducting phase existing, which is supported by XRD. But the transport measurements revealed that Pb addition only significantly improved the transport $J_c$ in low field region in the 122 tapes [62]. This is not surprised, since a remarkably enhanced magnetic $J_c$ in Pb-added samples originates from both intra-grain and inter-grain currents. As the SEM study on microstructures showed that Pb addition promotes crystal growth, these large grains, meaning large dimensions of intra-grain loops, were supposed to contribute to the enhancement of magnetic $J_c$ in the entire field region.

In conclusion, in the case of the 122 wires, the Pb addition can increase the transport $J_c$ in low field, but give no help in high field region. On the other hand, the Ag addition can effectively improve the transport $J_c$ in high field region. Thus, it is natural to consider that by combining the advantages of the doping effect of Ag and Pb, the $J_c$ performance can be enhanced in the whole field region. Recently, Yao et al. prepared Ba$_x$K$_{1-x}$Fe$_2$As$_2$ tapes containing Ag and Pb dopants, using the *ex situ* PIT method combined with a short high-temperature annealing technique [63]. Through co-doping with Ag and Pb, the transport critical current density of $Ba_xK_{1-x}Fe_2As_2$ tapes was significantly improved in whole field region and the highest transport $J_c$ was up to $1.4 \times 10^4$ A/cm$^2$ ($I_c$=100 A) at 4.2 K in 0 T, as shown in figure 16. It is believed that the superior $J_c$ values in the co-doped samples are due to the combined effects of Pb doping at low fields and Ag doping at high fields. However, the hysteretic effect of $J_c$ in increasing and decreasing fields suggests that there are still weak-linked current paths between grains in these tapes.

### 4.2 Thermal treatment

For the preparation of high quality 122 pnictide superconductors, Zhang et al. have attempted to systematically optimize sintering conditions [64]. First, the sintering time was fixed at 35 h, and the sintering temperature was varied from 700 to 1000$^o$C. Figure 17 shows X-ray diffraction patterns of all samples prepared at different temperatures. Powder XRD analyses revealed that the resulting bulks composed of almost a single phase of Sr-122, however, a small amount of FeAs as an impurity phase appears in the samples sintered at 700$^o$C. The content of FeAs is decreased by increasing the sintering temperature and tends to disappear at temperatures over 850$^o$C, indicating that the impurity phases can be effectively reduced by higher sintering temperatures, hence a well-developed 122 phase is formed.

As shown in figure 18, the $J_c$ at 20 K increases monotonically with increasing the sintering temperatures up to 850 ℃, then further increasing the sintering temperature hardly influenced the $J_c$ improvement. In addition, samples heat treated at temperatures over 850 ℃ exhibited a very weak $J_c$-field dependence. Characterizations revealed that high temperature sintering resulted in large grains with fewer impurities. Therefore the higher $J_c$ is mostly due to the improved connection of grains resulting from the decrease of impurity phases. Regarding the sintering temperature, it was found that over 850 ℃ is the optimized temperature for producing high $J_c$ 122 polycrystalline bulks.

However, in the case of Ag-added samples, the $J_c$ increases rapidly from 850 to 900ºC, and the $J_c$ value of Ag-doped samples finally exceeds the pure samples at 900ºC, as shown in the inset of figure 18. The result suggested that the effect of Ag on the $J_c$ property of Sr-122 is strongly correlated to the heat treatment temperature. Namely, the optimized sintering temperature for Ag-doped samples was found to be around 900 ºC [65], almost 50 ºC higher than the case of the bulks without Ag addition. This difference can be clarified from the XRD data (see figure 19). In addition to main 122 phases, some Ag and a small amount of FeAs and AgSrAs impurity phases were observed in all the samples heated at different temperatures. It is noted that sintering at low temperatures below 800 ºC, a large amount of AgSrAs phases were formed from the reaction of Ag, Sr and As, then these AgSrAs were gradually decomposed into Ag and 122 phase with increasing the sintering temperature. At 900 ºC, AgSrAs nearly disappeared in the samples, and again Ag appeared as the particle within the 122 phases, leading to the improved connectivity caused by either filling voids among grains by Ag or the decrease of impurity phases. The similar results were also observed in the Ag-added Ba-122 polycrystals [66]. These data indicate that the optimum sintering temperature could differ for pure and Ag-added 122 pnictides; for Ag-doped pnictides, a higher annealing temperature has proven to give better results.

Therefore, careful control of the sintering temperature can bring a significant reduction of nonsuperconducting phases, particular FeAs impurities, resulting in better grain connection and ultimately higher $J_c$.

### 4.3 Precursor powder

It is well known that controlling the properties of the precursor powder was vital to the PIT pnictide process and that any changes in the composition, or homogeneity of starting particles can affect the superconducting performance of pnictide wires. In particular, for synthesizing 122 polycrystalline samples from precursor powders, maintaining the initial content of K, which has high chemical reactivity with oxygen and high equilibrium vapour pressure at sintering temperatures, is very important. Any losses of K for forming the 122 phase would produce secondary phases because the 122 phase has none or quite small nonstoichiometry, resulting in poor reproducibility in the microstructures as well as the critical current properties.

Different approaches to prepare high quality precursor materials have been studied, for instance, we utilized a ball-milling technique to make a fine powder mixture of constituent elements, instead of hand grinding [44]. Togano et al. reported a new method to obtain high quality precursor powder, which starts from small pieces of Ba, K and commercially available FeAs alloy instead of powders and melting the mixture of pieces at a high temperature above the melting point [67]. In order to have a good mixing of constituent elements, the heat treatment was carried out at a high temperature above the melting point of the FeAs compound (~1050 °C) for a short time of 5-15 min. The XRD data shows that the obtained bulk material with the composition of (Ba, K)Fe$_2$As$_2$Ag$_{0.5}$ had strong peaks of 122 phases with fewer impurity phases, indicating that the reaction to form 122 phase completed in a short time in the molten state (see figure 20). The good formation of the 122 superconducting phase was also confirmed by the strong signal of diamagnetism, which is shown in the inset of figure 20. This method was effective to achieve a higher transport $J_c$ in the *ex situ* PIT process, because the sample had fewer impurity phases and a higher density compared to samples prepared by a conventional sintering process [68].

For Sr/Ba-122 compound, another challenge of having high quality polycrystalline samples is how to control the content of extremely active element K, even in a high-grade glove box. Like Mg in MgB$_2$, light element K has low melting point. As a result, burning loss of K is unavoidable, usually leading to K-deficient in 122 superconductor. Wang et al. found that over-doping of K in Ba-122 has been successful in raising $J_c$ [69]. Phase-pure polycrystalline Ba$_{0.6}$K$_{0.4+x}$Fe$_2$As$_2$ with $0 \leq x \leq 0.1$ were prepared using a one-step solid-state reaction method. As shown in figure 21, the $J_c$ shows strong dependence on K-doping level. $J_c$ increases with increasing K content and a maximum was observed for sample with $x = 0.1$. High-field $J_c$ for samples with $x = 0.1$ is three times higher than that for samples with $x = 0$, while over-doping of K has minimal effect on the $T_c$.

TEM study revealed that overdoped K samples had much higher concentration of dislocations than that in nominal samples, as shown in figure 22. Small distortions of lattice induced by over-doped K led to lattice dislocations, resulting in enhanced pining force and hence improved critical current density $J_c$. The present observation is in agreement with the Ref. [70] that K doping introduces K inhomogeneity, which results in electronic strain, acting as pinning centers. These results show that it is indispensable to prepare the 122 pnictides with appropriate K excess for obtaining excellent superconducting properties.

**4.4 Transport properties of 122 wires and tapes by the *ex situ* process**

As already mentioned, the early approach to fabricate the 122 wires was the *in situ* PIT technique [15, 44], however, the $J_c$ values were quite low, due to numerous non-superconducting phases, cracks and porosity, most of which were a consequence of the de-densification that occurred during the thermal process. Recent developments revealed that wires synthesized by the *ex situ* method have higher $J_c$ than the ones produced by the *in situ* technique, because the *ex situ* process has more options for remixing unreacted materials and reacting them multiple times to drive the reaction to completion, finally leading to fewer impurity phases as well as a high density of the superconducting core.

In early 2010, Qi et al. reported the first *ex situ* fabrication of 122 pnictide superconducting wires with an Fe/Ag double sheath and achieved significant transport critical currents up to 37.5 A at 4.2 K [71]. The $Sr_{0.6}K_{0.4}Fe_2As_2$ composite wires were prepared as followings: the raw materials (Sr filing, Fe powder, As and K pieces), were thoroughly grounded in Ar atmosphere for more than 5 hours using ball milling method. The powder was filled into a Nb tube, and then heated to 850$^o$C for 35 hours. The heated materials were added silver or lead powder (10 wt %, 200 mesh, 99.9% purity), and reground and filled into a bimetallic silver/iron tube of 10 mm outside diameter and 2 mm wall thickness. After packing, the tube was rotary swaged and then drawn to wires of 2 mm in diameter. The wires were cut into 8~10 cm, which were annealed at 900$^o$C for 20 hours.

Figure 23 presents the $J_c(H)$ values for wires based on the *ex situ* process. The wire with the addition Pb was found to exhibit the highest $J_c$ value at 4.2 K (3750

A/cm$^2$ at 0 T), a factor of ~12 higher than that of the pure wires. More importantly, a super current density of 100 A/cm$^2$ still flowed in the Ag-added wire even under high fields up to 8 T. In addition, the transport $J_c$-$H$ property in the higher K composition Sr$_{0.6}$K$_{0.5}$Fe$_2$As$_2$ tapes was further enhanced upon Pb doping [72], as shown in the inset of figure 23. Examination of the microstructure showed that the $J_c$ enhancement was due to the elimination of cracks and enhanced connectivity [71]. Thus, *ex situ* process seemed very effective for increasing $J_c$ in 122 pnictide wires.

In order to further improve the critical current properties, Togano et al. developed a melting process to prepare a high quality precursor material and successfully fabricated Ag-sheathed (Ba, K)Fe$_2$As$_2$+Ag superconducting wires by the *ex situ* PIT process [50]. The precursor material was prepared by a high-temperature synthesis above the melting point, which plus Ag addition were then ground into powder form. The powder was then put into a Ag pipe and the composite was cold worked into a wire and heat treated at 850$^o$C for sintering. In this case transport $J_c$ reached almost 10$^4$A/cm$^2$ at 0 T and 1000 A/cm$^2$ at 15 T at 4.2 K. Good grain connectivity in the precursor particles and good intergrain connection obtained by the Ag addition are considered to be responsible for this higher transport $J_c$. Recently, Ding et al. also reported Ag-sheathed (Ba, K)Fe$_2$As$_2$ wires with an improved transport $J_c$ of 1.3 ×10$^4$A/cm$^2$ at 4.2 K in 0 T with the help of Ag addition [51]. However, the application of much less than 1 T depresses $J_c$ by an order of magnitude, indicating the strong weak-linked behavior in their wires. This fact was further proven by the magneto-optical analyses.

Although many works can increase the $J_c$ in self-field and the $J_c$ plateau in the high-field region, they do not eliminate the rapid drop of $J_c$ in low field. Thus, besides reducing defects such as pores, cracks, and inhomogeneous phases in 122 tapes, solution of the weak-linked problem is another important issue to further improve the critical current.

**4.5 Grain boundary properties and texturing process**

As we know, a remarkable transport inter-grain $J_c$ of 10$^6$A/cm$^2$ at 4.2 K and 0 T has been reported in cobalt-doped BaFe$_2$As$_2$ films [54-55], offering the possibility of a number of practical applications. This is in contrast to polycrystalline samples, which show a relative low inter-grain (transport) $J_c$ of ~$10^4$ A/cm$^2$ at 4.2 K and 0 T, and a weak-linked behavior form a creep drop of $J_c$ at low field and a hysteretic phenomenon. The first critical step in understanding the mechanism by which polycrystalline samples can not support a high transport $J_c$ is to determine the structure and composition of the grain boundaries.

One way to directly observe structure, composition, and bonding effects at grain boundaries (GBs) is through the combination of high-resolution transmission electron microscopy (TEM) and electron energy loss spectroscopy (EELS) at atomic resolution in a scanning transmission electron microscope (STEM). Recently our TEM and EELS studies of nanoscale structure and chemical composition of grain boundaries in polycrystalline Sr-122 samples demonstrated that grain boundaries are usually coated by non-superconducting amorphous layers and there is significant oxygen enrichment in the amorphous layers at GBs [73], which are important for a better understanding of transport current in pnictide polycrystalline wires.

As shown in figure 24, three distinctive types of grain boundaries were found: clean grain boundaries, boundaries containing amorphous layer with a width of ~10 nm (i.e., larger than the coherence length), and boundaries containing amorphous-crystallite-amorphous trilayers ~30 nm in width, consequently resulting in a current blocking effect. Similar amorphous layers around individual grains have been reported in Sm- and Nd-1111 polycrystals [35, 74]. However, the EELS data (see figure 25) revealed the microscopic origin of the nanoscale non-stoichiometry at the grain boundaries, in particular, different levels of oxygen contamination, Sr redistribution and perhaps oxidation which may be responsible for these glassy layers. These amorphous layers at GBs would act as transport critical current barriers, and by itself may be enough to explain the relatively low transport $J_c$ of polycrystalline samples with respect to that of 122 pnictide films. It also suggests that reduction of the oxygen content during the fabrication process will be necessary to increase the intergrain $J_c$.

On the other hand, the properties of grain boundaries have been a critical issue for the application of superconducting materials, in particular, to wires and tapes [75-76]. For high-$T_c$ cuprates, the grains must be highly textured to prevent the deterioration of $J_c$ across misaligned GBs because the $J_c$ strongly depends on the misorientation angle of GBs. Regarding YBCO, $J_c$ across the grain boundary starts to decrease at a critical angle ($\theta_c$) of approximately 3-5$^o$ and shows nearly exponential rapid decay with further increasing the misorientation angle [76]. To reduce the weak-links, YBCO wires are produced by epitaxial growth on the biaxially aligned substrate or intermediate layer. On the other hand, for Bi2223 tapes by the PIT method, the mechanical processing is used to promote high grain alignment in order to bypass the problem of high-misorientation-angle grain boundaries.

Lee et al. were the first to study grain boundary properties of iron pnictide superconductors using Ba122:Co epitaxial films grown on [001]-tilt STO bicrystal substrates with four different GB misorientation $\theta$ of 3, 6, 9, and 24$^o$ [77]. Their essential conclusion is that $J_c$ sharply decreases as the misorientation angle increases from 3 to 24$^o$, and the critical angle is estimated to be 3-5$^o$ as shown in figure 26. they claimed that this weak-linked behavior is similar to those observed for YBCO.

Katase et al. performed a more systematic study on the transport properties of GBs with $\theta$=3-45$^o$ using Co-Ba122 epitaxial thin films with a self-field $J_c$ of above 1MA/cm$^2$ at 4 K that were directly prepared on both MgO and LSAT bicrystal substrates [78]. Contrary to the view of Lee et al. that the critical angle was 3-5$^o$ [77], they found that the critical angle for [001] tilt boundaries was ~9◦ (figure 27). This critical angle was substantially larger than the value of 3-5$^o$ reported for YBCO GBs [76]. It was then clear that the grain boundary suppression of $J_c$ is proportionately not quite as large as in YBCO, which allows us to use a simpler and lower cost production process to fabricate superconducting wires and tapes.

Although the grain boundaries in pnictides do not degrade the overall $J_c$ as heavily as YBCO, the $J_c$ values of wires only remain at ~$10^4$ A/cm$^2$ and the decay slopes with the field are large. These deteriorated properties may suffer from the existence of large angle GBs with $\theta_{GB}$ much greater than $\theta_c = 9^o$. As we know, an effective method to overcome the weak-linked problem is to engineering textured grains in iron pnictides to minimize deterioration of the $J_c$ across high-angle GBs. More recently we have developed a new method of deformation processing of *ex situ* PIT, iron-sheathed Sr-122 tapes to achieve *c*-axis aligned texture. This is followed by short high-temperature annealing to enhance the grain connectivity. Both XRD and SEM revealed that the presence of plate-like grains in preferred orientation. Indeed, the resultant Sr-122 tapes showed higher transport critical current densities than our previous wires. [79]

Immediately, we optimized the above texturing strategy by combining with the Sn addition and achieved a superior transport critical current of as high as 180 A in zero field at 4.2 K in $Sr_{0.6}K_{0.4}Fe_2As_2$ tapes [80]. The as-prepared samples were synthesized as followings: Mixtures of Sr filings, Fe powder, As and K pieces were ground using a ball milling method in Ar atmosphere for more than 10 hours, with the aim to achieve a uniform distribution. Raw powders were heat treated at 900°C for 35 h. For Sn doped samples, 10 wt% Sn was added to the precursor powder. The mixed precursors were ground and filled into an iron tube, which was subsequently swaged and drawn down to a wire of ~2.0 mm in diameter. The as-drawn wires were then cold rolled into tapes with a reduction rate of 10~20%. Fe is the suitable sheath material for the application of iron based superconductor. It is a cheap and abundant resource. The tapes were finally sintered at 1100°C for a short time of 1~15 minutes in Ar atmosphere.

SEM microstructures for fracture surface of Sr-122 tapes exhibit a dense layered structure, very similar to what has been observed in Bi2223 superconductors. EDX analysis on a large area in the Sn added samples clearly demonstrates that the sample is composed of Sr, K, Fe, As and Sn elements. As shown in figure 28, at 4.2 K the transport $J_c$ in self-field and 10 T exhibited values of $2.5 \times 10^4$ A/cm$^2$ and $3.5 \times 10^3$ A/cm$^2$, respectively, which is one of the highest $J_c$ performance reported to date [80]. These $J_c$ values of the tape samples highlight the importance of grain alignment and Sn addition for enhancing the $J_c$ of 122 superconducting wires. We should note that the grain alignment is still not perfect in the present sample, suggesting that a further enhancement in the $J_c$ performance can be expected upon the optimization of fabrication process and introduction of pinning centers.

The inset of figure 28 shows the field dependence of $J_c$ at 4.2 K of the typical tape sample for parallel and perpendicular applied fields. The in-field $J_c$ is slightly higher in the perpendicular field ($J_c^{\perp}$) than in the parallel field ($J_c^{//}$). This is the opposite of $MgB_2$ and Bi-2223, in which the $J_c^{//}$ is always larger than the $J_c^{\perp}$. However, the anisotropy ratio (~1.5) is not so large as in the case of Bi-2223 [81], but almost similar to that of $MgB_2$ [12].

Although the *ex situ* PIT process is very effective in enhancing $J_c$, the challenge of improving the phase purity of 122 wire is still significant. Recently, Weiss et al [52] have developed a technique to make high-purity Ba-122 Cu/Ag clad wires using the *ex situ* PIT technique combined with the hot isostatic press (HIP) under 192 MPa of pressure at 600 °C for 20 hours. The obtained wire claims up to $10^5$ A/cm$^2$ in 0 T at 4.2 K, further suggesting the important role of high quality precursors during the *ex situ* process. It is noteworthy that this process seems complicated, requiring three HIP heat treatments with an intermediate cold isostatic press (2 GPa) step. Most recently, we have fabricated even larger $J_c$ Sn-doped Sr122 tapes by optimized a heat treatment process plus texturing process, which are both effective ways to improve grain connectivity. The most striking and important feature of our data is the magnetic field dependence which is extremely weak, for instance, the $J_c$ of these tapes exhibits above $10^4$ A/cm$^2$ in a high magnetic field of 10 T (see figure 29), indicating a promising future for high-field applications. The preparation and details of superconducting properties of these high performance Sr-122 tapes will be reported elsewhere [82].

So far, the variation of $J_c$-H at 4.2 K for 122 type tapes and wires by the *ex situ* PIT process is summarized in figure 29. Critical current densities of the order of $10^4$-$10^5$ A/cm$^2$ have been reported. These data clearly demonstrate that significant progress in raising the transport $J_c$ of wires is occurring and PIT approaches may actually be amongst the most promising to apply for making wire conductors for practical applications.

**5. 1111 type wires and tapes**

The 1111 phase materials have short coherence length [17], low carrier density

[83], significant evidence for granularity and low intergranular $J_c$ [35, 36], which are quite similar to the case of cuprates. Global supercurrent flow [36, 84], and locally well-connected area [85, 86] were confirmed to exist in this system.

Although 1111 type superconducting wires were synthesized in 2008 [13, 14, 87], no transport critical current was observed due to strong reaction occurred between the superconducting core and the sheath, since the 1111 wires were heat treated at temperatures as high as ~1200°C at that time. Subsequently, Wang et al. have systematically studied the influence of sintering temperature on the superconductivity of polycrystalline $SmFeAsO_{0.8}F_{0.2}$ [88], and found that samples sintered at a low temperature clearly show high $T_c$ as shown in figure 30, for example the $T_c$ of the samples sintered at 850°C is even above 53 K, and the samples prepared at 1000°C display the highest $T_c$ of 56.1 K reported so far. Furthermore, the samples sintered at 900-1000°C show the higher RRR and the lower $\rho$ (57 K), indicating the low impurity scattering and enhanced carrier density. These results suggest that annealing at a temperature of below 1000°C seems also suitable for obtaining high quality 1111 phase oxypnictides, compared to the commonly used temperatures of around 1200°C.

Based on this finding, we successfully fabricated Sm-1111 wires by the *in situ* PIT process at a temperature as low as 900°C [46], about 300 K lower than those reported previously. In order to avoid the reaction problem, silver was employed as a sheath material, like the case of 122 wires. Most importantly, as the red circle line in figure 31 shows, it was the first 1111 type wire that carried transport supercurrent, attaining a $J_c$ of 1300 A/cm$^2$ at 4.2 K, 0 T, which almost kept constant in a field of over 0.5 T. Meanwhile, we have made superconducting Sm-1111 tapes with a silver sheath and demonstrated a transport $J_c$ of as high as 2700 A/cm$^2$ (4.2 K, 0 T) [72]. The onset $T_c$ occurs at 45.2 K, and zero resistance is achieved at 40.5 K. Note that the tape samples showed higher $J_c$ than the wire samples. This is reasonable because it can be attributed to the increased packing density during the rolling of the wire into a flat tape.

Higher sample density may increase the number of local links and strong links, leading to better connectivity thus more intergranular current loops. Kursumovic et al.

reported that very higher density Nd-1111 polycrystals were made by spark plasma sintering method, giving density values above 95% of the theoretical value while maintaining very high slopes of upper critical field of up to 8.5 $T/K$ [89]. Ding et al. [90] prepared a series of polycrystalline Sm-1111 with different sample densities, and found that the superconducting volume fraction, the critical current density and even the pinning force density were improved with increasing the sample density due to the improvement of local links. However, Yamamoto et al. reported that even for the purest polycrystalline 1111 bulk samples made by the sintering and HIP process, their intergranular current densities are still small [84], providing further evidence that polycrystalline 1111 oxypnictides, like cuprates, are largely affected by weak-linked problems.

As mentioned early, *ex situ* process is a good mean for remixing unreacted materials and reacting them multiple times to get high purity phase. Recently, Fujioka et al. reported the *ex situ* PIT fabrication of superconducting Sm-1111 wire by using an Ag sheath [91]. In order to compensate for the F losses, a binder with stoichiometric Sm, Fe, As, and F was added during the second sintering process, as a result, a $J_c$ of 4000 A/cm$^2$ at 4.2 K, 0 T was obtained. Shortly after that, Wang et al. improved the starting materials and succeeded in fabricating Ag-sheathed Sm-1111 tape samples with less arsenide impurities. A transport $J_c$ was further increased to 4600 A/cm$^2$ at 4.2 K and 0 T [92].

In fact, some progresses have been made in the 1111 wire fabrication as shown in figure 31, however, the in-field $J_c$ values are still quite low, e.g., more than one order of magnitude lower at a field of over 1 T, compared to those of 122 wires, suggesting the difficulties in wire fabrication and fluorine controlling. Future studies may focus on the methods of texturing and element-doping in order to minimize weak-linked problems at grain boundaries and to improve the intergranular current density of polycrystalline 1111 superconducting wires.

## 6. 11 type wires and tapes

The FeSe (11 type) compounds have the lowest $T_c$ among the iron-based superconductors. However, their advantages of simplest crystal structure and containing no toxic arsenic lead to many studies for both fundamental physics and potential applications.

Mizuguchi et al. reported the fabrication of the first 11 type compound wire, an Fe(Se,Te) wire with a $J_c$ of approximately 12.4 A/cm$^2$ (4.2 K, 0 T) by using the modified powder-in-tube method, an Fe sheath, and TeSe powder. [93] A better result has been obtained by the same group in the FeTe$_{0.5}$Se$_{0.5}$ wires by the *ex situ* PIT process [94], where $J_c$ reaches 64 A/cm$^2$ at 4.2 K in 0 T. They suggested that the $J_c$ enhancement was related to improvement of the intergrain connectivity by heat treatment.

It is clear that the transport $J_c$ of 11 wire fabricated by PIT method is very low, because it is difficult to achieve high-density 11 wire using the PIT process due to the package or the shrinkage of core materials. In order to get much denser and more homogeneous superconducting core, Gao et al. developed a novel gas diffusion process for preparing high quality FeSe superconductor wires and tapes [95]. The diffusion technique only requires a shorter reaction time than conventional sintering processes.

Figure 32*c* shows the final product of wires and tapes with an FeSe layer ranging from several microns to about 0.1 mm thick, depending on the treatment process and the amount of Se powder. The obtained FeSe samples, with a very dense structure with few cracks (see figure 32 *a* and *b*), had an onset temperature of about 15.1 K, and zero resistivity was attained below 9.3 K, which is comparable to the sample value prepared by the solid state reaction method [96]. Interestingly, an enhancement of $J_c$ up to 137 A/cm$^2$ at 4 K and self-field was observed. Subsequently, upon optimizing the diffusion process, Ding et al. further achieved a $J_c$ as high as 600 A/cm$^2$ (4.2 K, 0 T), and also found a distribution of $T_c$ and weak-linked features by magneto-optical imaging [97]. Recently, Takano et al. also fabricated three- or seven-cores 11 wires by a Fe-diffusion process. The obtained $J_c$ (0 T) is 588 A/cm$^2$ for the three-core wire and 1027 A/cm$^2$ for the seven-core wire, but wires became one or two orders of magnitude lower when the field was applied. [98, 99] These results demonstrate that the diffusion process is one of the most effective ways to realize higher $J_c$ in 11

superconducting wires.

More recently, Palenzona et al. reported a more significant enhancement $J_c$, reaching about $10^3$ A/cm$^2$ (4.2 K, 0 T) especially with the very weak field dependence behavior in a polycrystalline 11 bulk synthesized by a melting process and a subsequent annealing treatment [100], however, no wire data was available in this literature. The $J_c$-$H$ characteristics of 11 wires fabricated by different methods are shown in figure 33, clearly, attempts to improve the transport $J_c$ in 11 wires have been less successful. Therefore, the $J_c$ values obtained are still one to two orders of magnitude lower than for the 122 wires, suggesting that there is more room for improvement.

## 7. Iron pnictide coated conductors

To date, 11, 122 and 1111 superconducting thin films on different single substrates have been fabricated, mostly by pulsed laser deposition (PLD) or by molecular beam epitaxy (MBE). [9, 54, 101-104] At 4.2 K, superior $J_c$ of several MA/cm$^2$ in 0 T has been reported. On the other hand, the evidence for weak-linked behavior from the 122 thin film bicrystals [77, 78] have naturally generated interest in preparing pnictide films on the textured metal substrates with biaxially textured buffer layers, which have led to high $J_c$ for YBCO coated conductors.

By employing an iron buffer layer architecture, Iida et al. realized the biaxially textured growth of Co-Ba122 thin films on IBAD-MgO-buffered Hastelloy substrates [105], which are typically used for the fabrication of coated conductors. The films exhibited in-plane misorientation angle of about 5.1 °, which was slightly smaller than that of the homoepitaxial MgO/IBAD-MgO layer. A self-field $J_c$ of over $10^5$ A/cm$^2$ has been achieved even at 8 K. Very soon, Katase et al. directly prepared biaxially textured Co-Ba122 thin films on IBAD-MgO buffered Hastelloy substrates [106]. The IBAD-MgO substrates consisted of a homoepitaxial MgO layer/IBAD-MgO layer/Y$_2$O$_3$ buffer layer/Hastelloy C276 polycrystalline tape. Such films exhibited high self-field $J_c$ values of 1.2-3.6MA/cm$^2$ at 2 K. Si et al. reported the fabrication of Fe(Se,Te) thin films on IBAD-MgO-buffered Hastelloy substrates.[107] They observed a nearly isotropic $J_c$ of over $10^4$ A/cm$^2$ at 4.2 K under a magnetic field as high as 25 T, indicating the potential of the iron-based superconductor for future high field applications.

However, the coated conductor technology, which has been developed for the second-generation YBCO films, has the shortcoming of low production rate and complexity, in particular, the high production cost disadvantages of textured template processing, elaborate vacuum systems required for physical vapor deposition, used in most buffer and epitaxial superconducting layer growth. Furthermore, it is hard to be applied to the volatile elements in iron based superconductors such as alkali metals doped 122 and F doped 1111 phases. In contrast, in the past two years, the $J_c$ has been rapidly enhanced for pnictide wires and tapes fabricated by the low-cost powder-in-tube process, e.g., the $J_c$ values have reached over $10^4$A/cm$^2$ in a field of 10 T at 4.2 K. As the cumulative knowledge of pnictide PIT process grows, it is believed that the critical currents will continue to increase. The potential of 122 pnictide wires and tapes can be estimated from figure 34, which shows the variation of $J_c$-$H$ for 122 PIT tape, Co-doped Ba-122 film, conventional Nb-Ti and Nb$_3$Sn wires, and PIT MgB$_2$ wire. In particular, the pnictide wire exhibits extremely weak field dependence in high field region, suggesting very promising for high field application at around 20 K, where Nb-Ti and Nb$_3$Sn wires can never be used owing to their $T_c$ below 20 K and MgB$_2$ wire is also difficult to be employed because of a very rapid drop of $J_c$ with the applied field [68].

Compared to film technique, the well-developed powder-in-tube method has great advantages of low cost and easiness of scale-up production. We believe that PIT process can be applied industrially to fabricate pnictide wires and tapes, as already demonstrated by the production of Bi-2223 and MgB$_2$ wires.

## 8. Conclusions and future outlook

More than four years have elapsed since the discovery of iron-based superconductor. The combination of extremely high $H_{c2}$ values and smaller anisotropy makes this class of superconductors appealing for high-field applications [108, 109]. As the first step to realizing applications, the fabrication of wires and tapes by the PIT method has been studied. Up to date, the field of iron-based wire superconductivity has seen a great progress from theoretical as well as the experimental point of view. Herein we reviewed the recent advances of iron-based wire research, covering the aspects of fabrication, characterization, and some remarkable technical breakthroughs. For convenience, we summarize Table II with the wires' relevant properties, including type, critical temperature and critical current density, whenever appropriate for easy reference and comparison. The current status of wire research area is summarized as follows:

1) Wires and tapes of 1111, 122 and 11 type compounds have been fabricated by the PIT methods, however, 122 wires showed the highest $J_c$ among the Fe-based superconducting wires, reaching above $10^4$ A/cm$^2$ at 10 T at 4.2 K, which are at least two orders of magnitude higher than for the 11 and 1111 wires

2) Ag was proved to be the best sheath material preventing from the reaction with pnictide. But recently Fe appeared as an alternative too, if the final heat treatment time was very short.

3) Compared to the *in situ* process, *ex situ* PIT process was more effective to reduce impurity phases, leading to substantial enhancements of $J_c$ for pnictide wires.

4) Addition such as Ag or Sn, substantially enhanced the high-field capability of iron pnictide wires, while Pb addition improved the transport $J_c$ in low field region.

5) For 122 pnictides, amorphous layers with oxygen enrichment at the grain boundaries were found to be the primary factor limiting fabrication of high $J_c$ polycrystalline conductors. On the other hand, overdoping of potassium can effectively enhance the $J_c$ due to the increased flux pinning.

6) The new PIT texturing process, using powers of Sn-added compounds, yielded 122 tapes with excellent in-field $J_c$ performance. Superior $J_c$ was attributed to well aligned superconducting grains and strengthened intergrain coupling achieved by Sn addition.

As already described, the powder-in-tube technique has been rather progressing as an alternative technology for iron-based superconducting wires and tapes, however, there are some points that are addressed for future prospect:

1) Phase purity of the wire samples is an important factor, e.g., for the 122 compounds, the K is highly volatile and has strong affinity to oxygen during the fabrication, K loss and formation of oxygen-rich amorphous layers are the main causes for the inhomogeneities and impurities. Understanding and reducing non-superconducting phases is an urgent subject in order to exclude the extrinsic factors. An expectation for significant $J_c$ enhancement depends on new techniques for controlling them.

2) Another problem to the PIT method is related to a densification of the filament, as only dense core yields improved $J_c$ values, because voids and impurities reduce the connectivity between grains, squeezing and lengthening the percolating current paths. More works are needed to decrease the porosity and to increase the density of the pnictide wires.

3) Since the pnictides are still weak-linked, wires need to use the same texturing strategies that have been so crucial for Bi2223 cuprate, as already demonstrated by the recent result [80]. Therefore, more efforts are required to understand how the texture is formed during the PIT deformation process of drawing, rolling and subsequent heat treatment and how it affects the transport properties.

4) Attention is also paid to the possibility that the $J_c$ will be raised by working effectively by addition technique.

5) The fabrication of long length pnictide wires by the PIT method seems relatively easy and more economical, however, there are no reports on this issue, due to short period after the discovery of pnictides. For practical applications, the next goal will be the development of long lengths of multifilamentary iron-based conductors with high $J_c(H)$ values. Like the case of $MgB_2$ and Bi2223 conductors, there are many critical issues, which need to be addressed seriously: i) Phase purity and homogeneity, which are very important issues for developing long length wires. ii) The fabrication of long length conductors at a reduced cost and the improvement in the performance of the conductors. iii) High thermal stability and mechanical properties, by optimizing deformation sequences yielding finer filaments and/or by using complex sheath geometries containing a Cu stabilizer.

In a word, it is very promising that the powder-in-tube method can be applied industrially to fabricate pnictide wires in km length.

Given the long period over which all other superconductors have been developed to the present state, we believe that challenges toward the realization of high performance pnictide wires and tapes will lead to an important breakout of these materials to practical applications.

**Acknowledgments**

I would like to take this opportunity to thank my colleagues Xianping Zhang, Zhaoshun Gao, Dongliang Wang, and my students Lei Wang, Yanpeng Qi, Zhiyu Zhang, Chunlei Wang, Chao Yao and He Lin for their work on pnictides. I wish to thank our collaborators, whose published works are quoted in this paper. The author is also indebted to Prof. H. Hosono for his great encouragement and useful comments. This work is partially supported by the National '973' Program (Grant No. 2011CBA00105), National Natural Science Foundation of China (Grant No. 51025726) and the Beijing Municipal Science and Technology Commission under Grant No. Z09010300820907.

Table I   Physical properties of iron-based superconductors in comparison with those of cuprates and MgB$_2$

|  | **Iron pnictides** | **Cuprates** | **MgB$_2$** |
|---|---|---|---|
| Parent materials | Antiferromagnetic metal (semi-metal) | Mott insulator | Metal |
| Fermi level | 5 bands | Single band | 2 bands |
| Pairing symmetry | s-wave | d-wave | s-wave |
| Impurity | robust | sensitive | sensitive |
| Max $T_c$ | 55 K | 134 K | 39 K |
| $H_{c2}$(0 K) | 100-250 T | ~100 T | 40 T |
| γ | 1-2 (122) | 5-7 (YBCO); 50-90 (Bi-based) | ~3.5 |

Table II Summary of transport $J_c$ properties at 4.2 K for different iron-based superconducting wires and tapes.

| Phase | Composition: dopant | Sheath | Synthesis condition | Type | $T_c$ (K) | $J_c$ at 0T (A/cm$^2$) | $J_c$ at 10T (A/cm$^2$) | Ref. |
|---|---|---|---|---|---|---|---|---|
| **1111** | | | | | | | | |
| | LaFeAsO$_{0.9}$F$_{0.1}$ | Fe/Ti | In situ, 1150°C/40h | Wire | ~25 K | -- | -- | Gao et al [13] |
| | SmFeAsO$_{1-x}$F$_x$ | Ta | In situ, 1180°C/45h | Wire | 52 K | -- | -- | Gao et al [14] |
| | SmFeAsO$_{0.8}$F$_{0.2}$ | Ta | In situ, 1150°C/48h | Wire | 52.5 K | -- | -- | Chen et al [87] |
| | SmFeAsO$_{0.7}$F$_{0.3}$ | Ag | In situ, 900°C/40 h | Wire | 41 K | 1300 | -- | Wang et al [46] |
| | SmFeAsO$_{0.8}$F$_{0.2}$ | Ag | In situ, 900°C/40 h | Tape | 45.2K | 2700 | -- | Ma et al [72] |
| | SmFeAsOF | Ag | Ex situ, 900°C/4 h | Wire | 54 K | 4000 | -- | Fujioka et al [91] |
| | SmFeAsOF | Ag | In situ, 900°C/30 h | Tape | 47 K | 4600 | -- | Wang et al [92] |
| **122** | | | | | | | | |
| | Sr$_{0.6}$K$_{0.4}$Fe$_2$As$_2$ | Nb | In situ, 850°C/35 h | Wire | 32.7 K | -- | -- | Qi et al [15] |
| | Sr$_{0.6}$K$_{0.4}$Fe$_2$As$_2$: Ag | Ag/Fe | In situ, 900°C/35 h | Wire and tape | 34 K | 1200 | 100 | Wang et al [44] |
| | Sr$_{0.6}$K$_{0.4}$Fe$_2$As$_2$: Pb | Ag/Fe | In situ | Tape | -- | 1100 | -- | Wang et al [62] |
| | Sr$_{0.6}$K$_{0.4}$Fe$_2$As$_2$: Ag or Pb | Ag/Fe | Ex situ, 900°C/20 h | Wire | 35 K | 3750 | 130 | Qi et al [71] |
| | (Ba, K)Fe$_2$As$_2$: Ag | Ag | Ex situ, 850°C/30 h | Wire | 35 K | $1.01 \times 10^4$ | 1100 | Togano et al [50] |
| | Ba$_{0.6}$K$_{0.4}$Fe$_2$As$_2$: Ag | Ag | Ex situ, 600-900°C/12-36 h | Wire | 37 K | $1.3 \times 10^4$ | -- | Ding et al [51] |
| | (Ba, K)Fe$_2$As$_2$: Ag+Pb | Fe | Ex situ, 1100°C/5 min | Tape | 33.7 K | $1.4 \times 10^4$ | -- | Yao et al [63] |
| | Sr$_{0.6}$K$_{0.4}$Fe$_2$As$_2$:Pb | Fe | Ex situ, texturing | Tape | 32 K | 5400 | -- | Wang et al [79] |
| | Sr$_{0.6}$K$_{0.4}$Fe$_2$As$_2$:Sn | Fe | Ex situ, texturing | Tape | 36.5 K | $2.5 \times 10^4$ | 3500 | Gao et al [80] |
| | (Ba$_{0.6}$K$_{0.4}$)Fe$_2$As$_2$ | Ag/Cu | Ex situ, HIP/600°C/20 h | Wire | ~36 K | $10^5$ | ~8500 | Weiss et al [52] |
| | Sr$_{0.6}$K$_{0.4}$Fe$_2$As$_2$:Sn | Fe | Ex situ, texturing | Tape | ~36 K | -- | $>1.5 \times 10^4$ | Gao et al [82] |
| **11** | | | | | | | | |
| | Fe(Se, Te) | Fe | Ex situ, 500°C/2h | Tape | 11 K | 12 | -- | Mizuguchi et al [93] |
| | FeTe$_{0.5}$Se$_{0.5}$ | Fe | Ex situ, 500°C/2h | Wire | 15.7 K | 64 | -- | Ozaki et al [94] |
| | FeSe | Fe | Diffusion, 400-800C | Wire and tape | 15.1 K | 137 | -- | Gao et al [95] |
| | FeSe | Fe | Diffusion, 600-800°C | Tape | 8 K | 600 | -- | Ding et al [97] |
| | FeSe | Fe | Ex situ, 1000°C/5h | 3 core wire | 10 K | 588 | ~17 | Mizuguchi et al [98] |
| | FeSe | Fe | In situ, 800°C/2h | 7 core wire | 10.5 K | 1027 | ~28.5 | Ozaki et al [99] |

**Figure 1** Crystal structure of the four categories of iron pnictides (a) '1111' type, (b) '122' type, (c) '111' type and (d) '11' type.

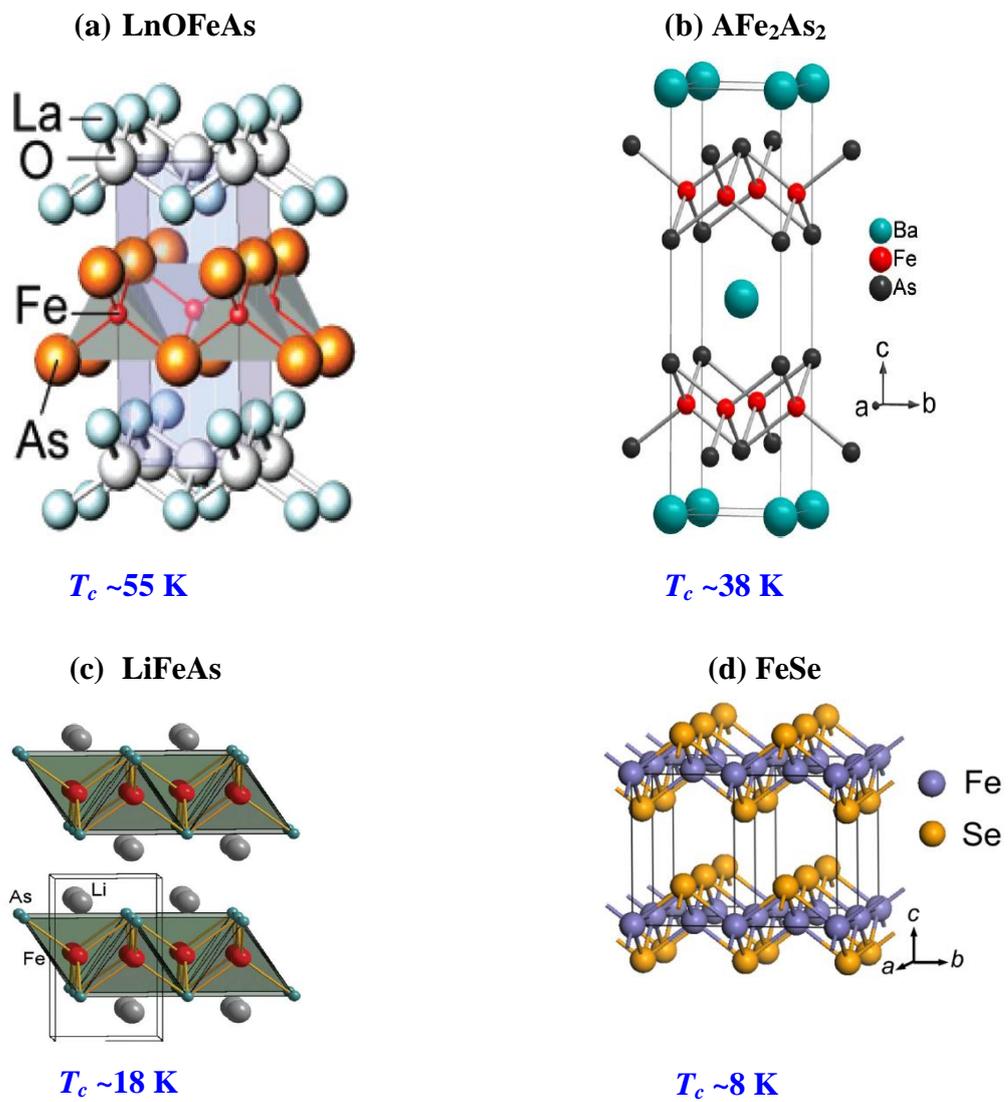

**Figure 2** Comparative *T-H* phase diagram for different superconducting materials [30]. Here the solid and dashed lines show the upper critical fields $H_{c2}$ and the irreversibility fields $H^*$, respectively.

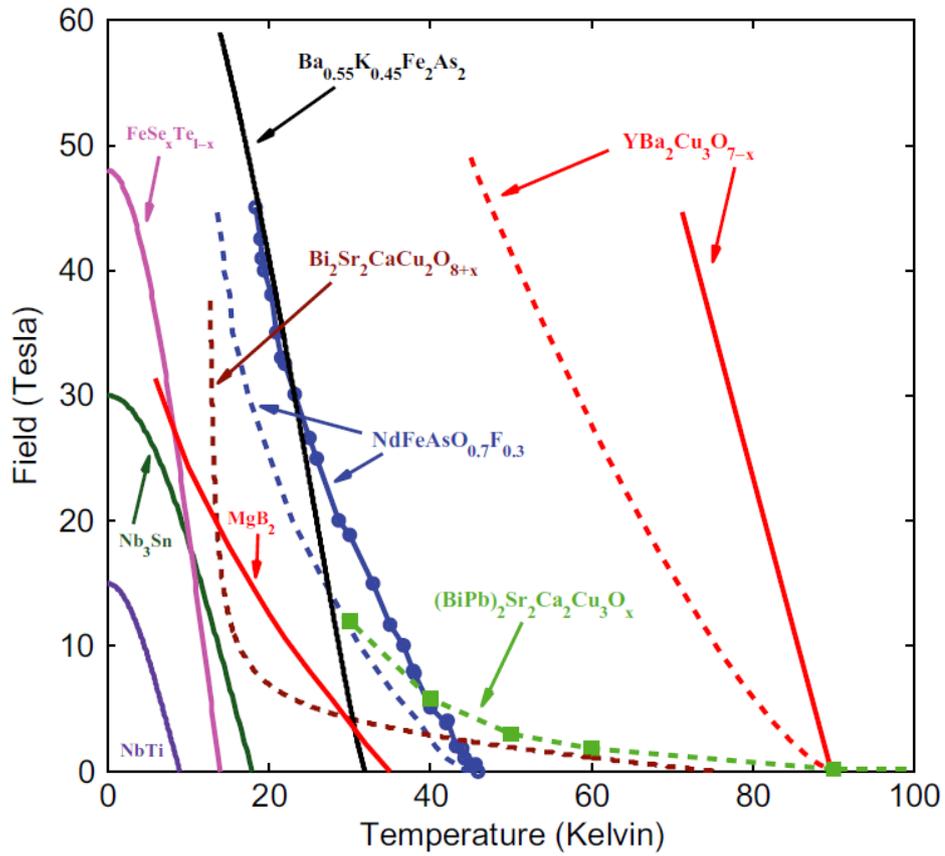

**Figure 3** Powder-in-tube (PIT) process used for fabricating iron pnictide wire. The main difference between *in situ* and *ex situ* processing, resides in the initial powder mixture.

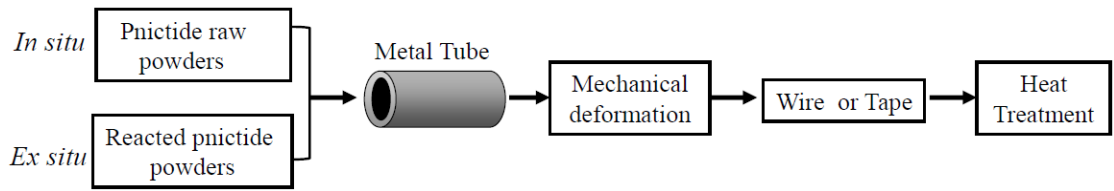

**Figure 4** Photograph of the SmFeAsO$_{1-x}$F$_x$ wires. SEM images for a typical transverse (a) and a longitudinal (b) cross-section of the wire after heat treatment [10].

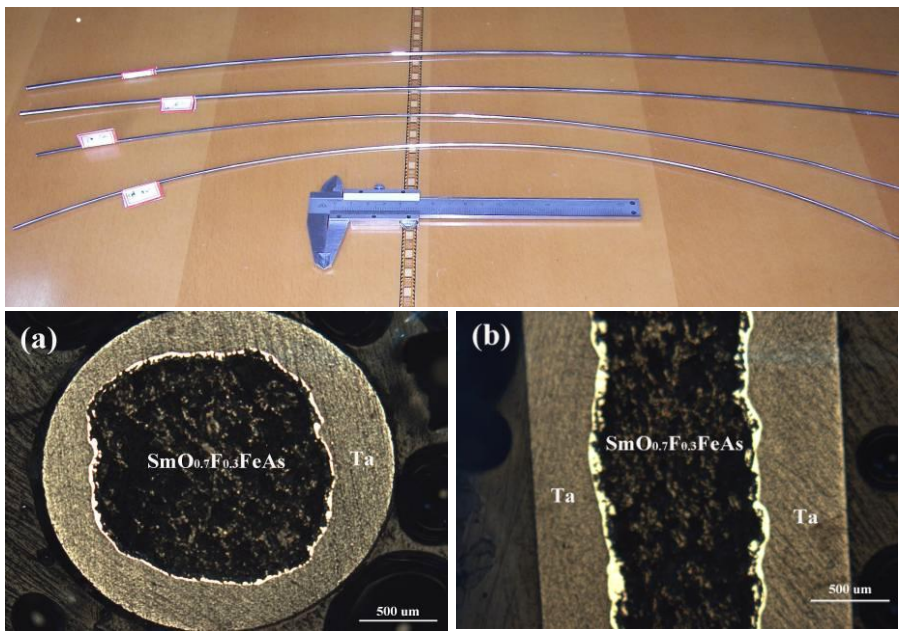

**Figure 5** Temperature dependences of resistivity for $SmO_{1-x}F_xFeAs$ filaments after peeling away the Ta sheath [14].

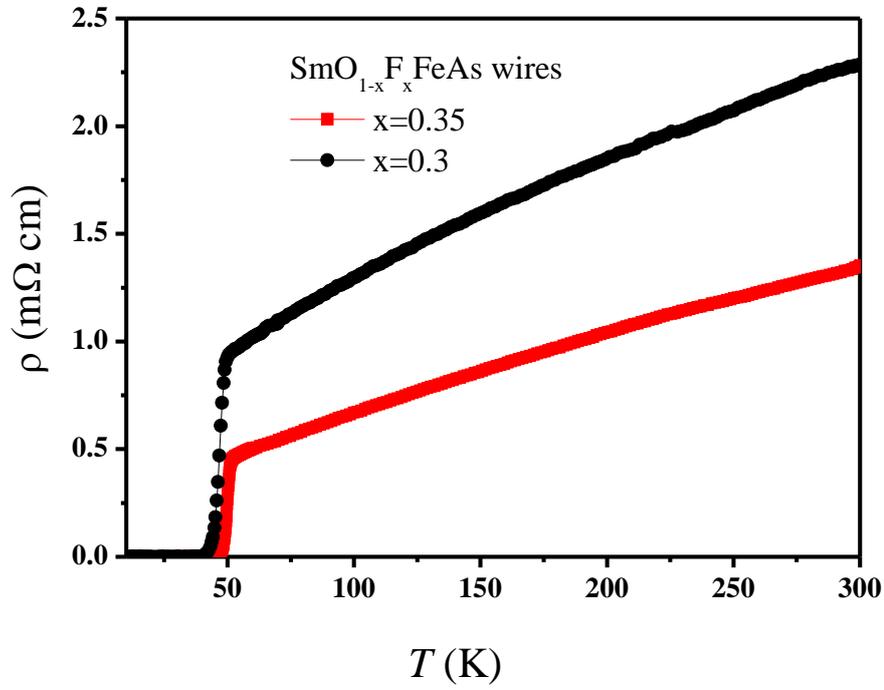

**Figure 6** Magnetic field dependence of $J_c$ at 5 K for the bar and powder of $SmO_{1-x}F_xFeAs$ samples [14].

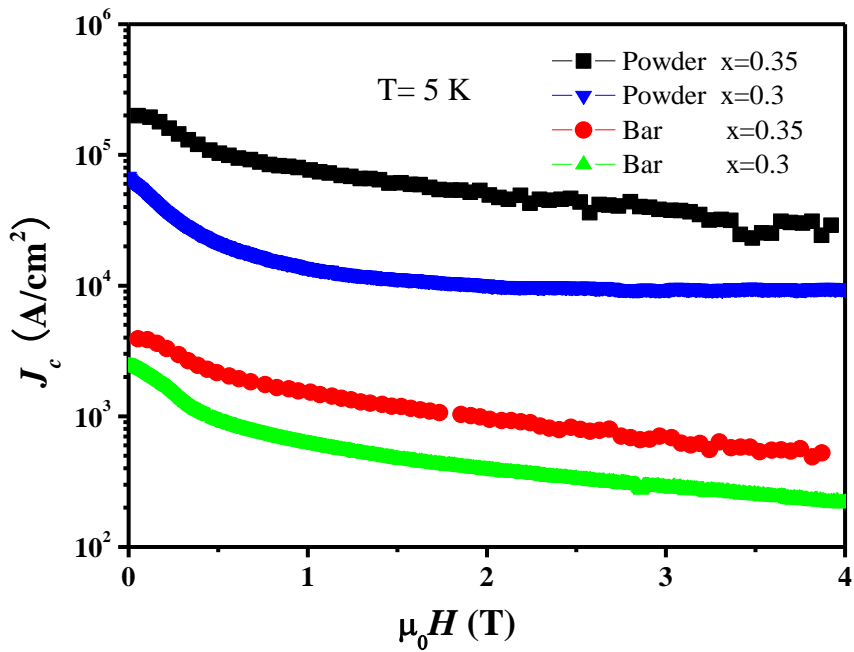

**Figure 7** Magnetic field dependence of $J_c$ at different temperatures for $Sr_{0.6}K_{0.4}Fe_2As_2$ wires [15].

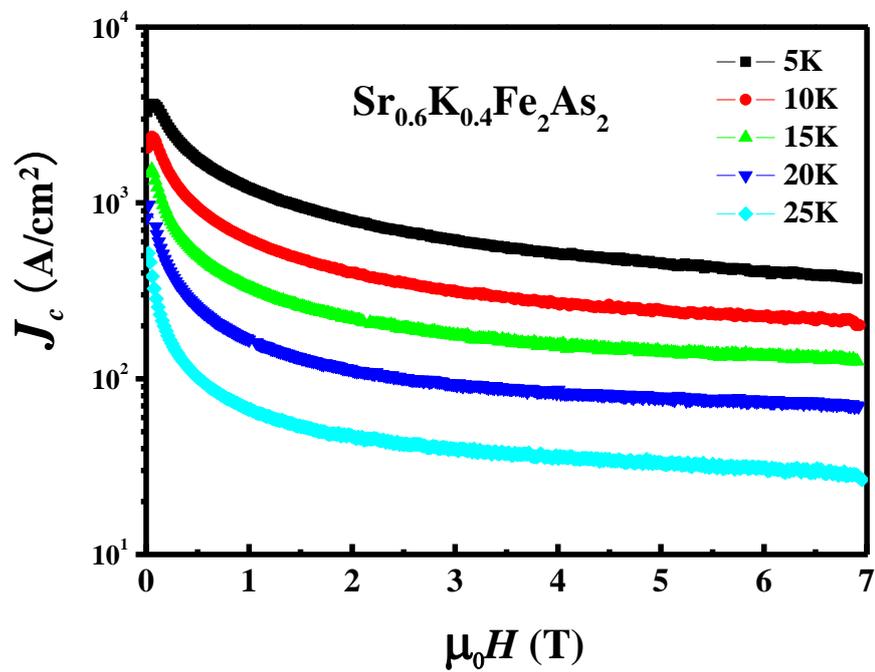

**Figure 8** (a) Transverse cross-sections of the typical Fe/Ag/Sr-122 wire and tape taken after heat treatment. (b) Magnified optical image of the Ag/ Sr-122 interface. [44]

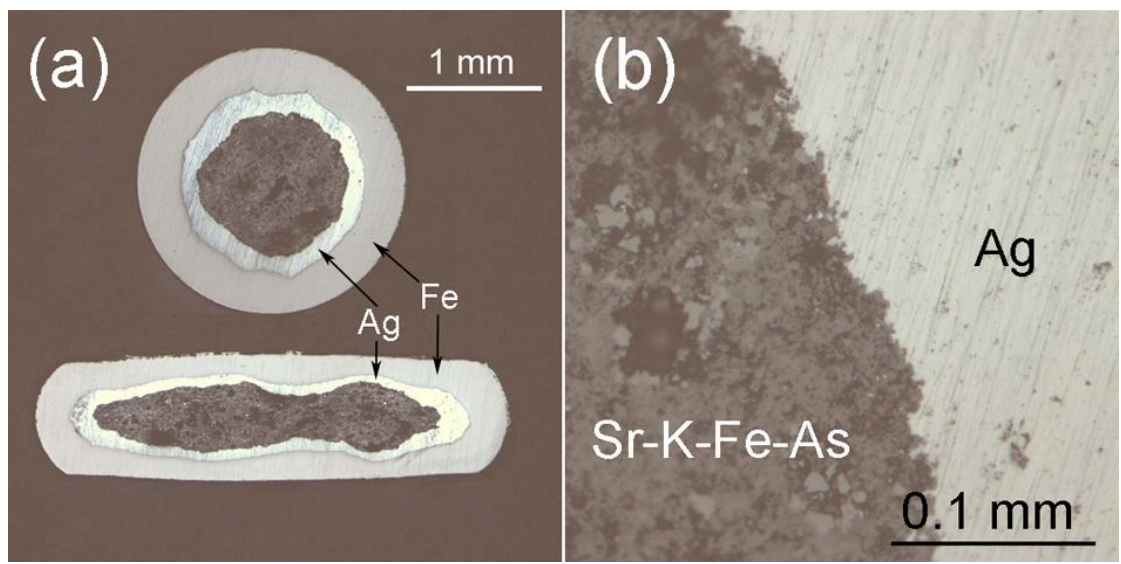

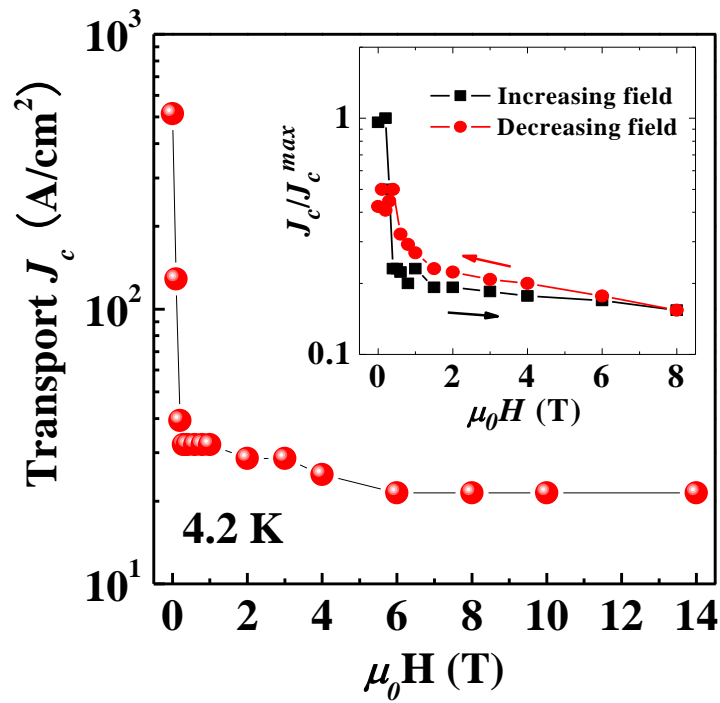

**Figure 9** Magnetic field dependence of transport $J_c$ for the first Ag-sheathed Sr-122 tapes. Inset: hysteresis in a normalized $J_c$.

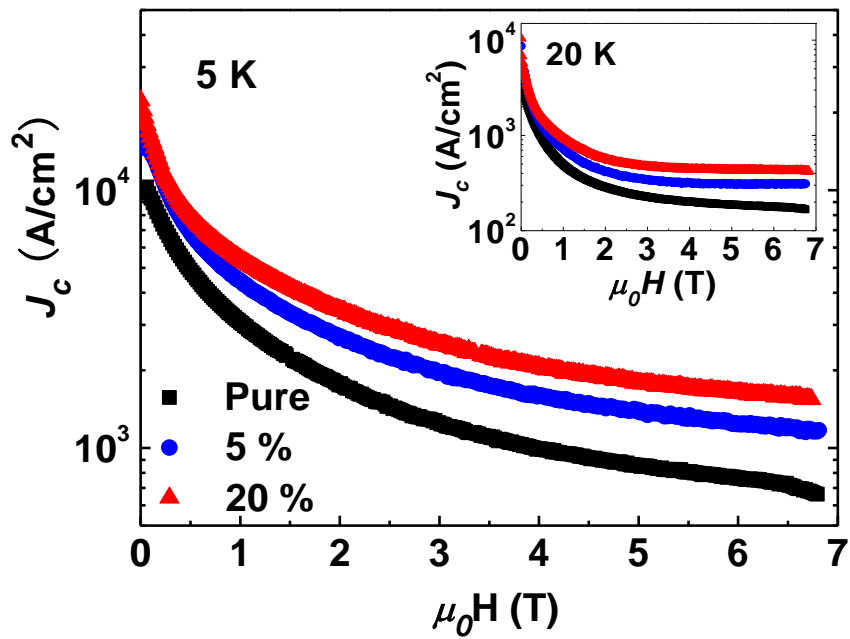

**Figure 10** Critical current densities derived from the magnetic hysteresis for various Ag-added samples at 5 and 20 K (inset) [60].

**Figure 11** Scanning electron images of the polished surface of the pure (a, b) and Ag-added (c, d) Sr-122 polycrystals.

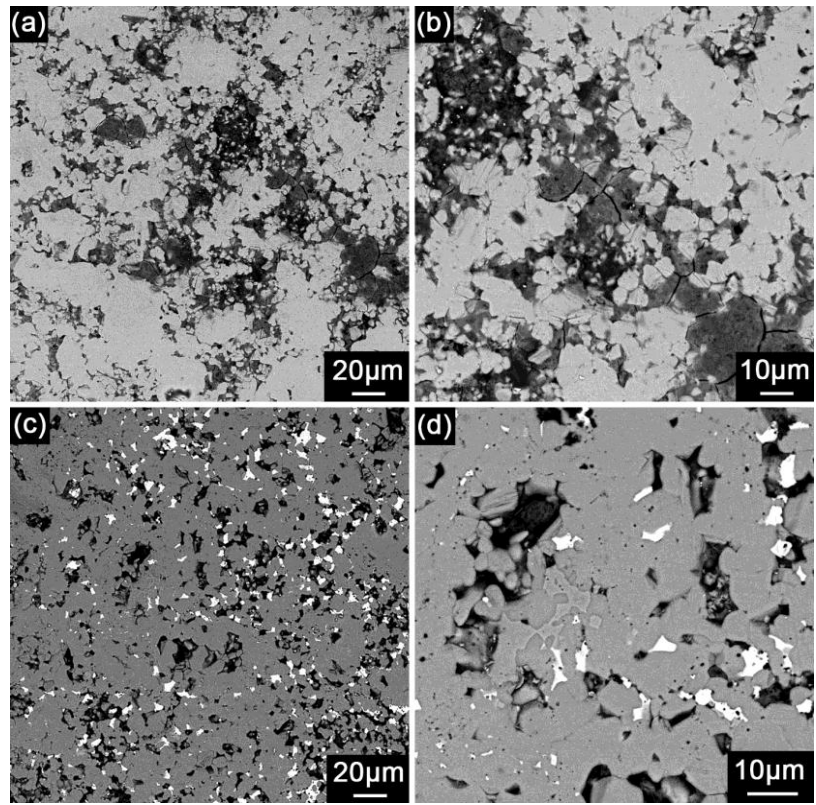

**Figure 12** Derivative of remanent magnetic moment as a function of maximum external magnetic field at various temperatures for pure and Ag-added samples [61].

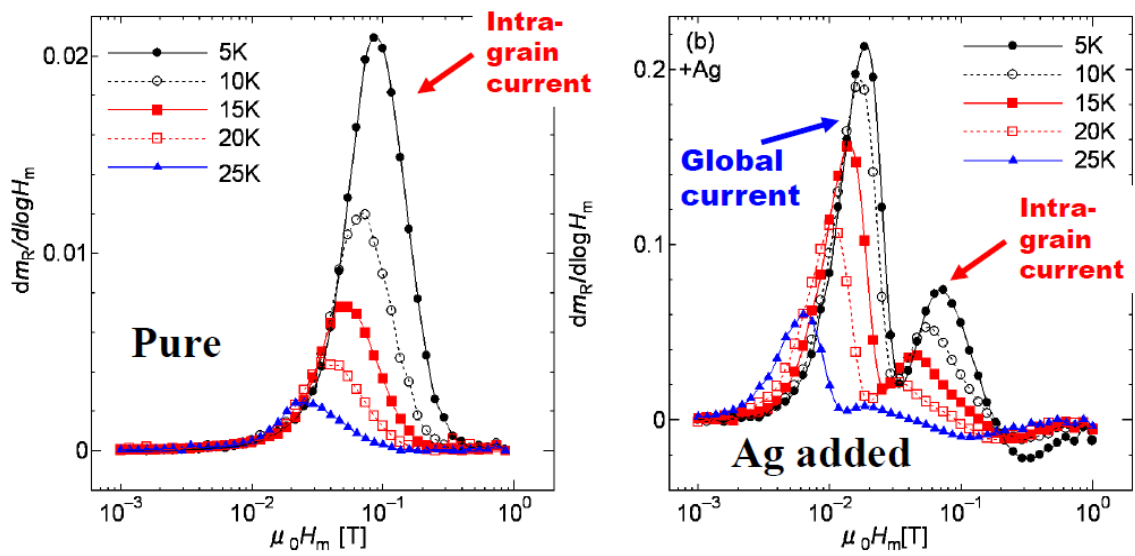

**Figure 13** Transport $J_c$ values of pure and Ag-added 122 tapes

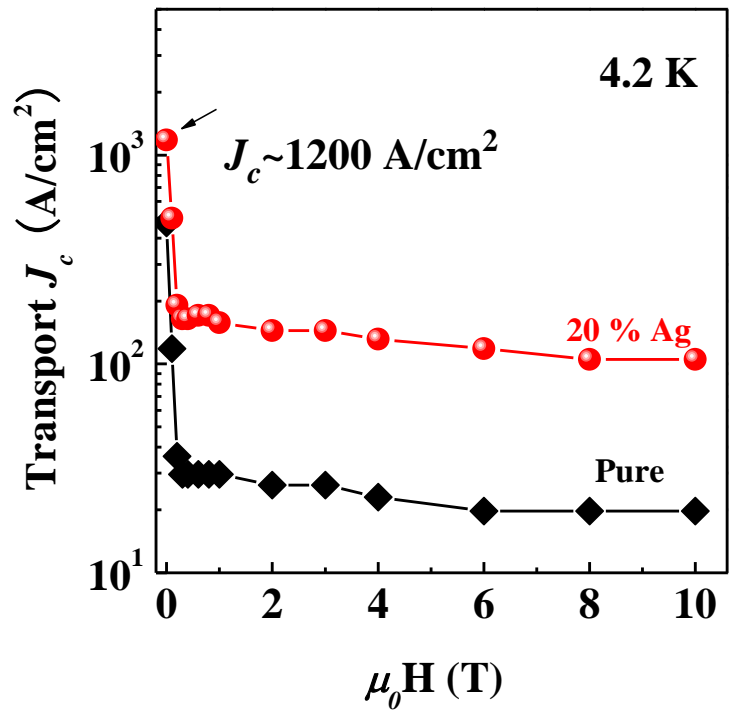

**Figure 14** SEM micrographs of superconducting cores of the pure (a and c) and Ag-added (b and d) 122 tapes [44].

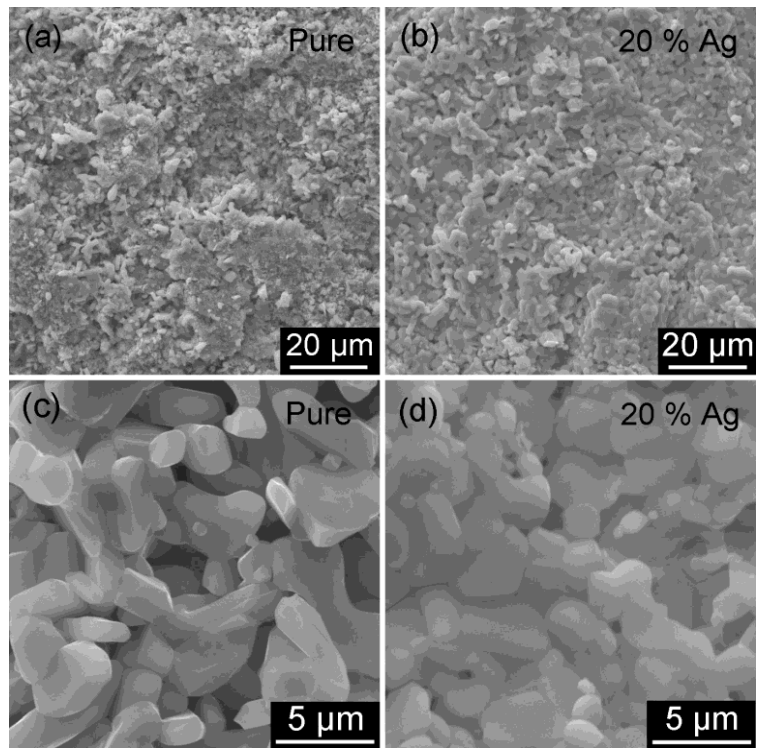

**Figure 15** Variation of magnetic $J_c$ as a function of applied field for pure and Pb added $Sr_{0.6}K_{0.4}Fe_2As_2$ bulk samples at 5 K; Inset: The magnetic $J_c$ of pure and 5 % Pb added $Sr_{0.6}K_{0.4}Fe_2As_2$ samples at various temperatures. [62]

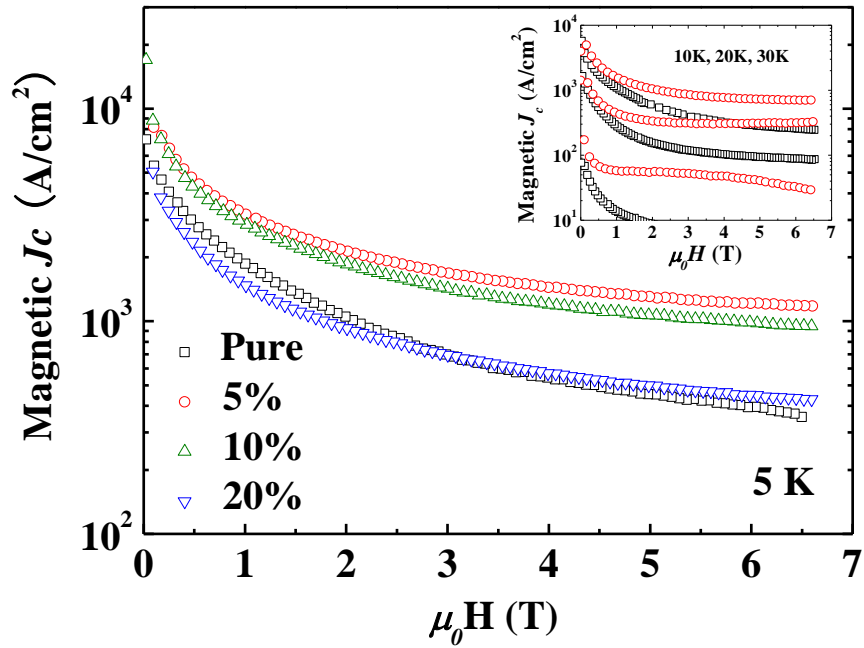

**Figure 16** Magnetic field dependence of transport $J_c$ at 4.2 K for Pb-doped and Ag+Pb-doped Ba-122 tapes. The inset shows the $J_c$ of AgPb-5 tape measured in increasing and decreasing fields successively. [63]

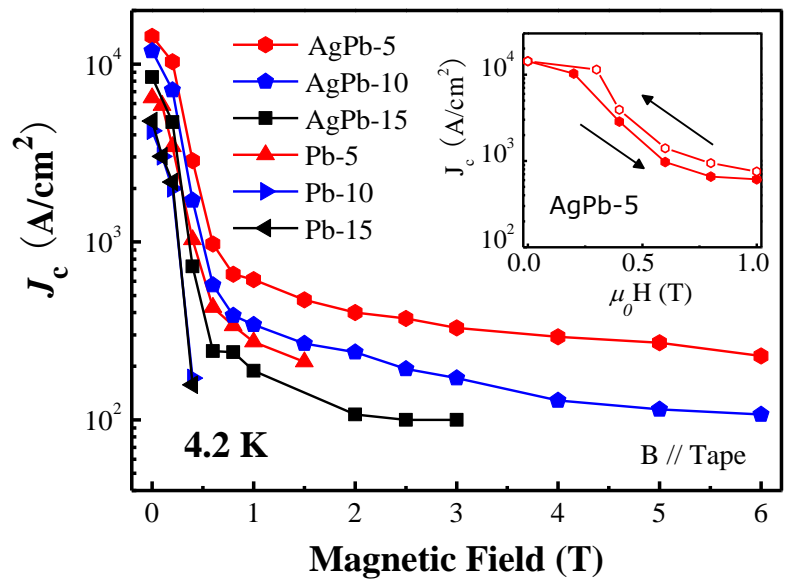

**Figure 17** X-ray diffraction patterns for Sr-122 samples at different annealing temperature [64].

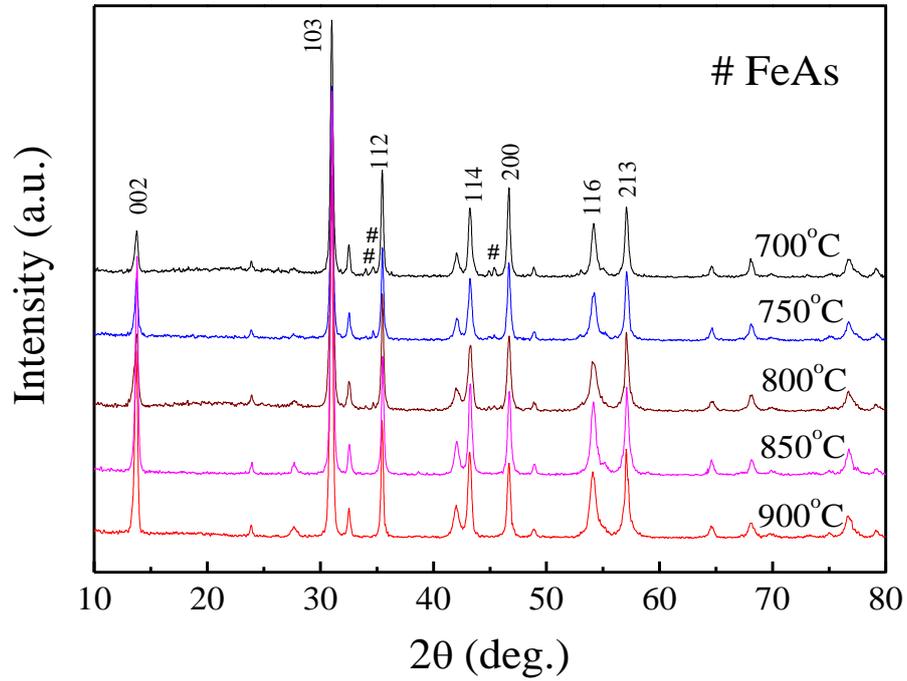

**Figure 18** Magnetic field dependence of $J_c$ at 20 K for the Sr-122 samples at different annealing temperature. The inset shows dependence of $J_c$ at 20 K and 6 T on the sintering temperature for pure and Ag doped Sr-122 samples.

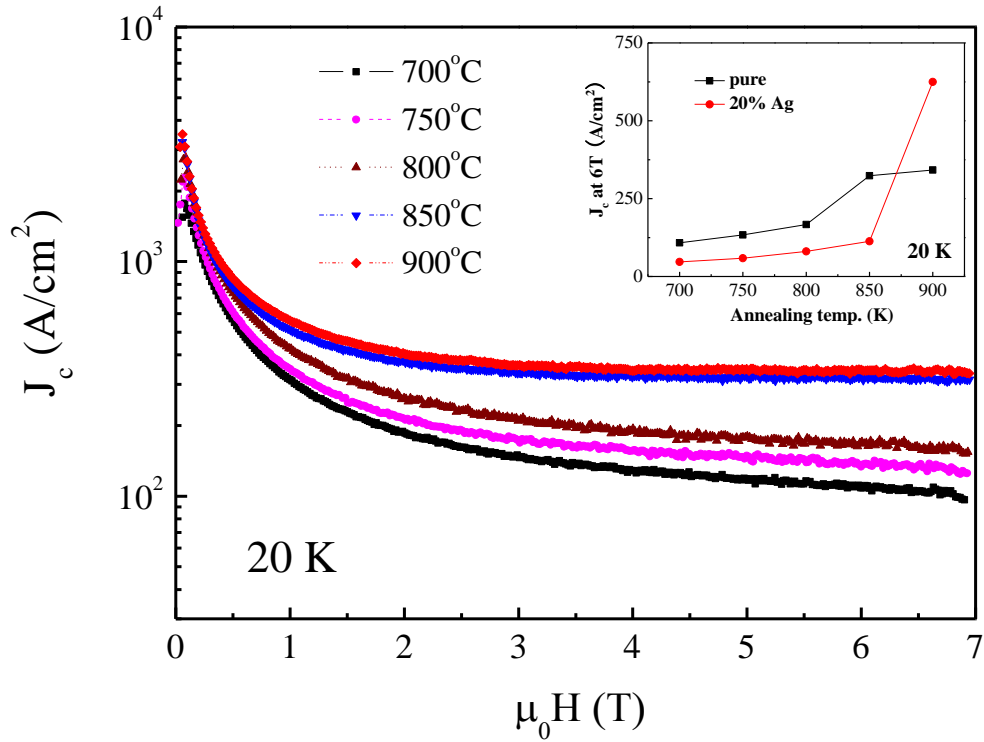

**Figure 19** X-ray diffraction patterns of the Ag-doped Sr-122 samples processed at different annealing temperatures [65].

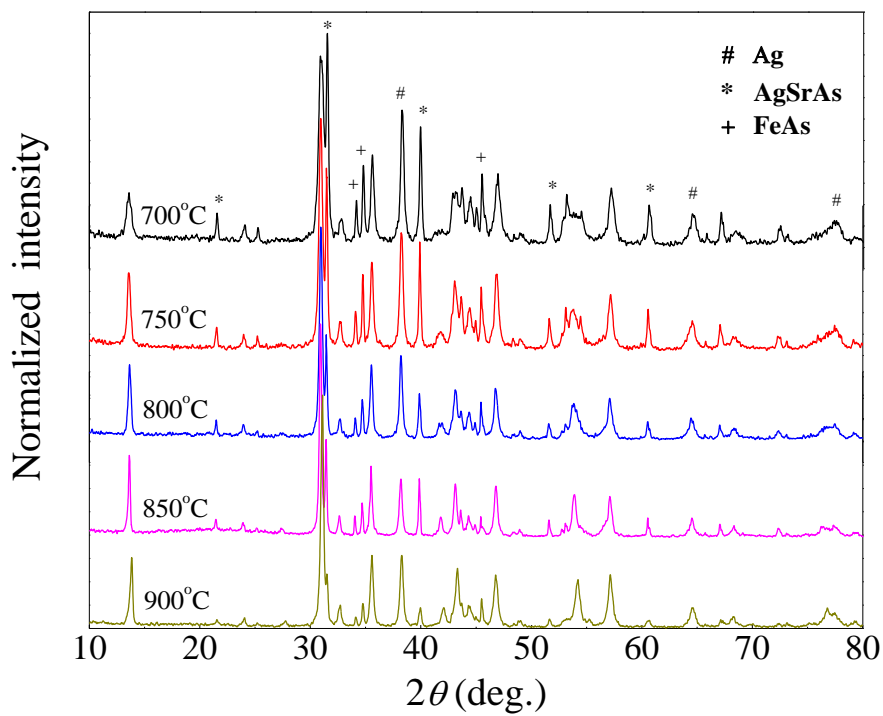

**Figure 20** Powder x-ray diffraction pattern of the melt processed (Ba, K)Fe$_2$As$_2$Ag$_{0.5}$ bulk material. The inset is the magnetization vs. temperature curve. [68]

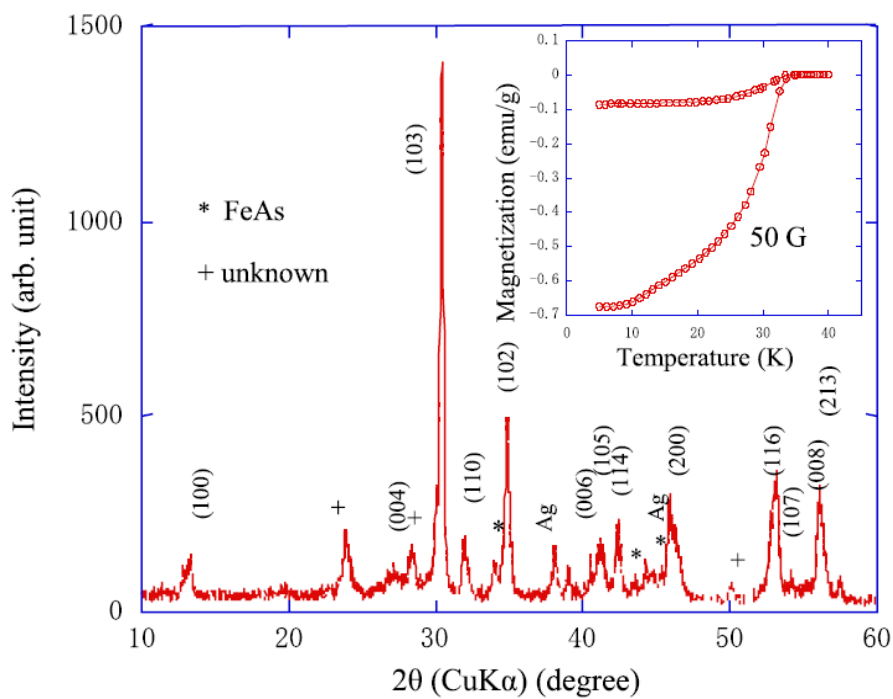

**Figure 21** Magnetic field dependences of $J_c$ at 5 K for bulk samples with various K-doping levels. Inset: Magnetic field dependences of $J_c$ at 5 K for powder samples with $x = 0$ and $x = 0.1$. [69]

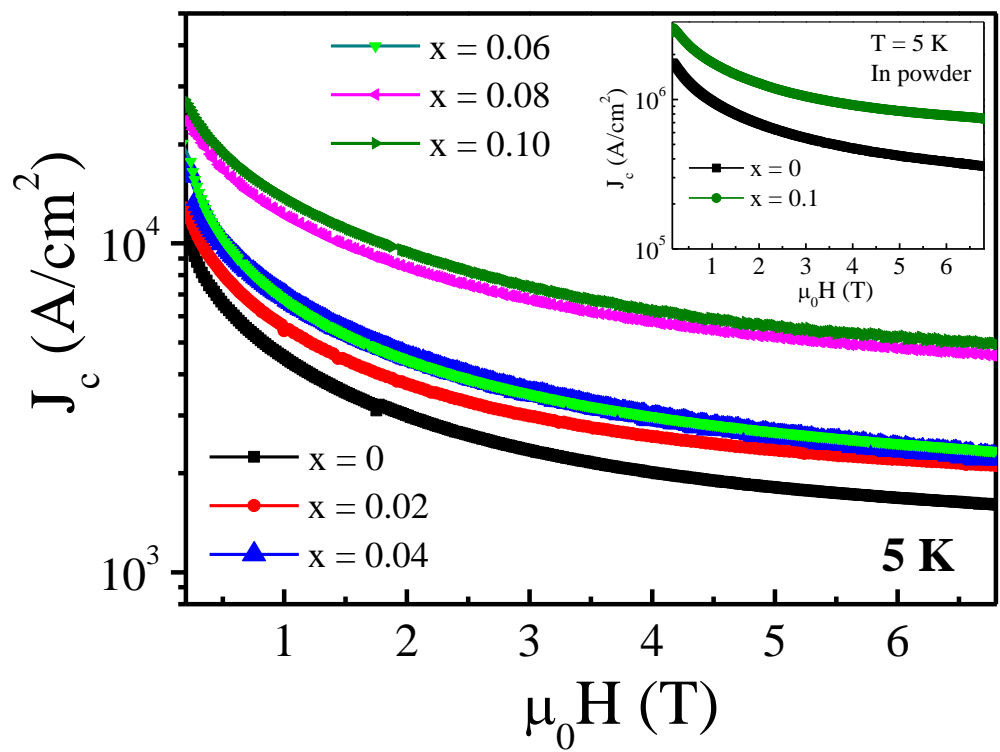

**Figure 22** TEM images for the $Ba_{0.6}K_{0.4+x}Fe_2As_2$ samples with (a) $x = 0$ and (b) $x = 0.1$ [69].

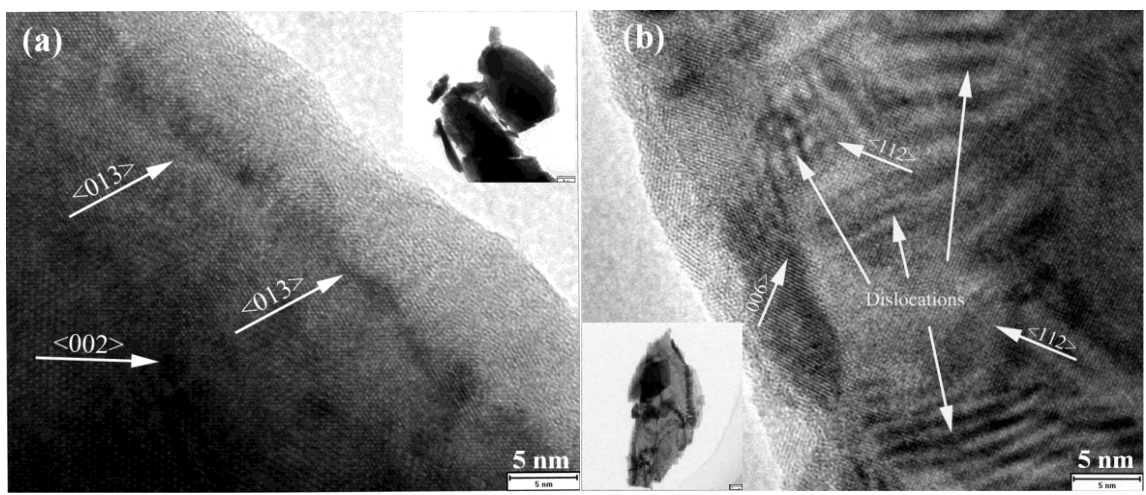

**Figure 23** Transport $J_c$ as a function of applied field for Ag-sheathed pure, Ag- and Pb-doped $Sr_{0.6}K_{0.4}Fe_2As_2$ wires. The inset shows transport $J_c$-$H$ curves of pure and Pb-doped $Sr_{0.6}K_{0.5}Fe_2As_2$ tapes. [72]

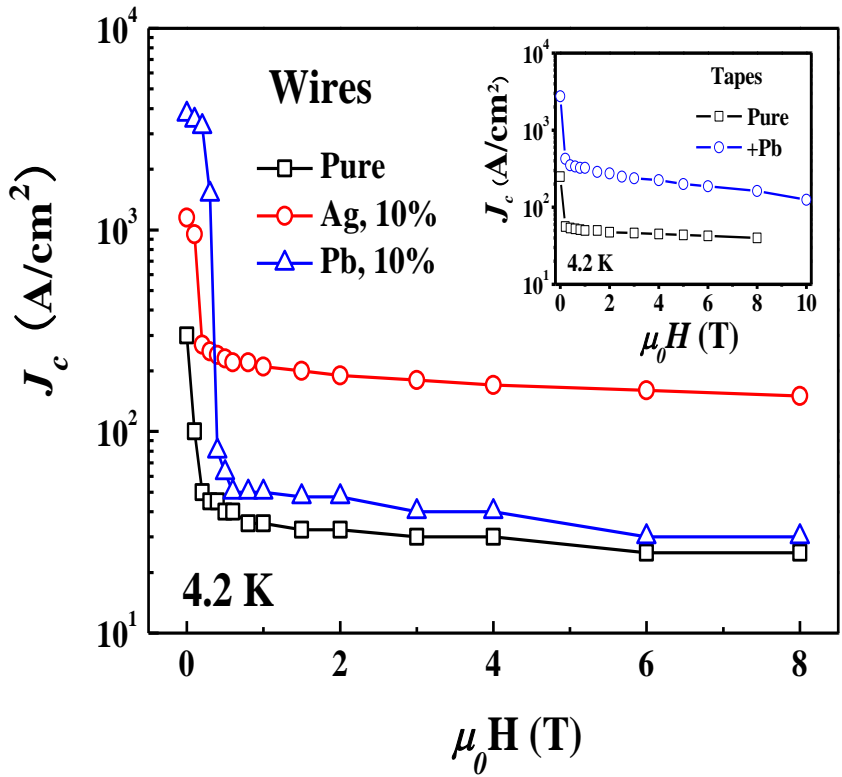

**Figure 24** (a) TEM image of an apparently well connected grain boundary network in polycrystalline $Sr_{0.6}K_{0.4}Fe_2As_2$. (b) A high-resolution TEM image of a typical, clean high-angle grain boundary. (c) The detailed structure of a grain boundary containing an amorphous layer about 10 nm in thickness. (d) Another type of grain boundary containing nanometer-scale impurity crystallites sandwiched between amorphous layers. [73]

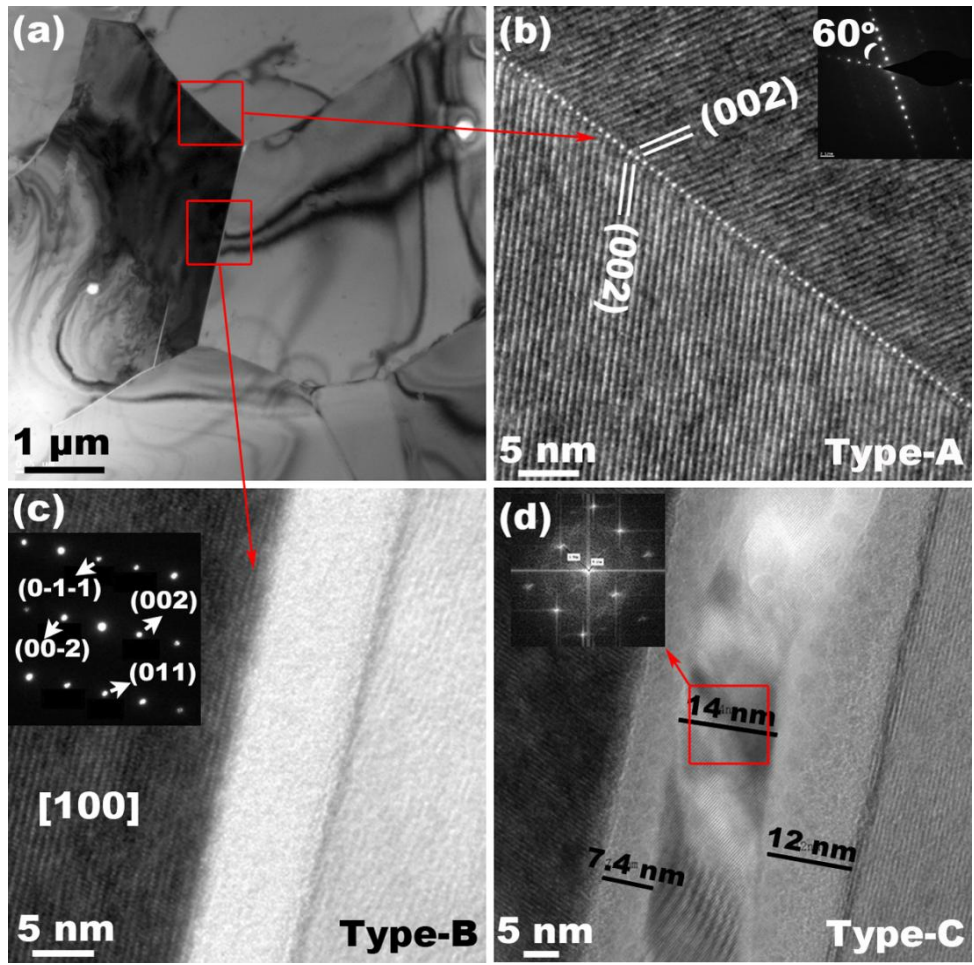

**Figure 25** (a) A STEM image of the Type-B grain boundary. (b) HAADF-line scan across the grain boundary. (c, d) EDS spectra and line scan performed on the amorphous layer. (e, f) EELS spectra and line scan performed on the amorphous layer. [73]

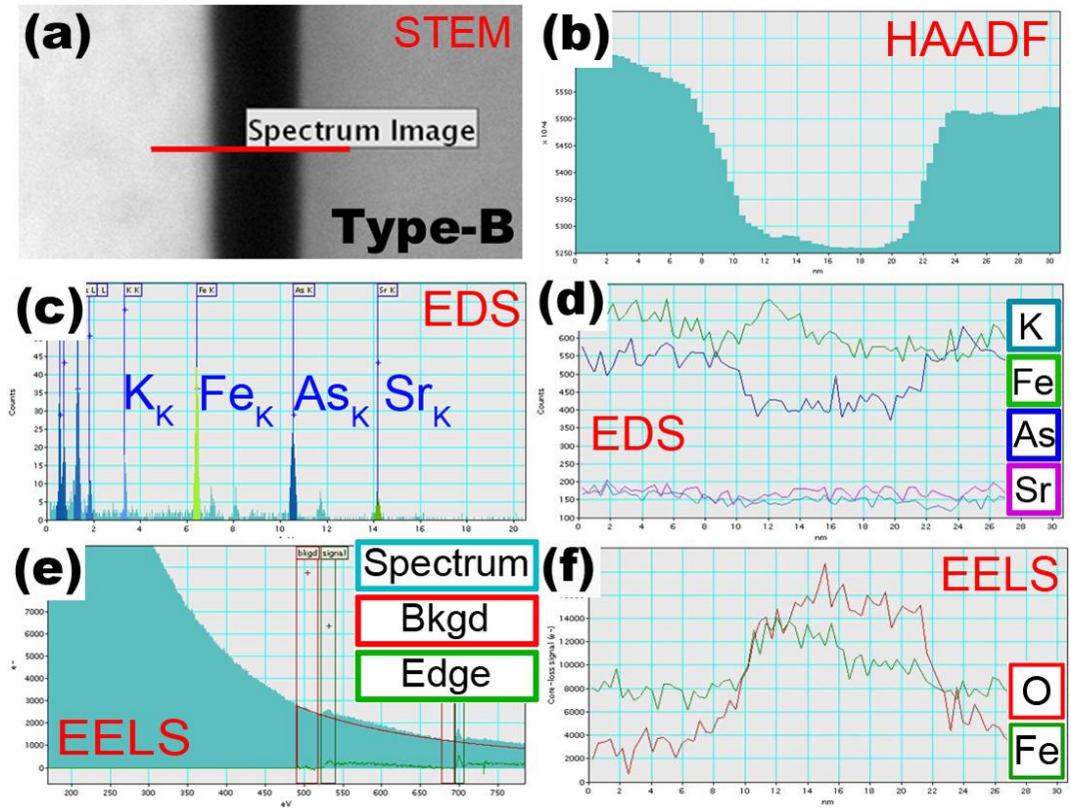

**Figure 26** Misorientation angle of GB dependence of $J_c$ of Ba122:Co on [001]-tilt STO bicrystal substrates at 12 K under an external magnetic field of 0.5 T. The inset shows the corresponding results for YBCO. [77]

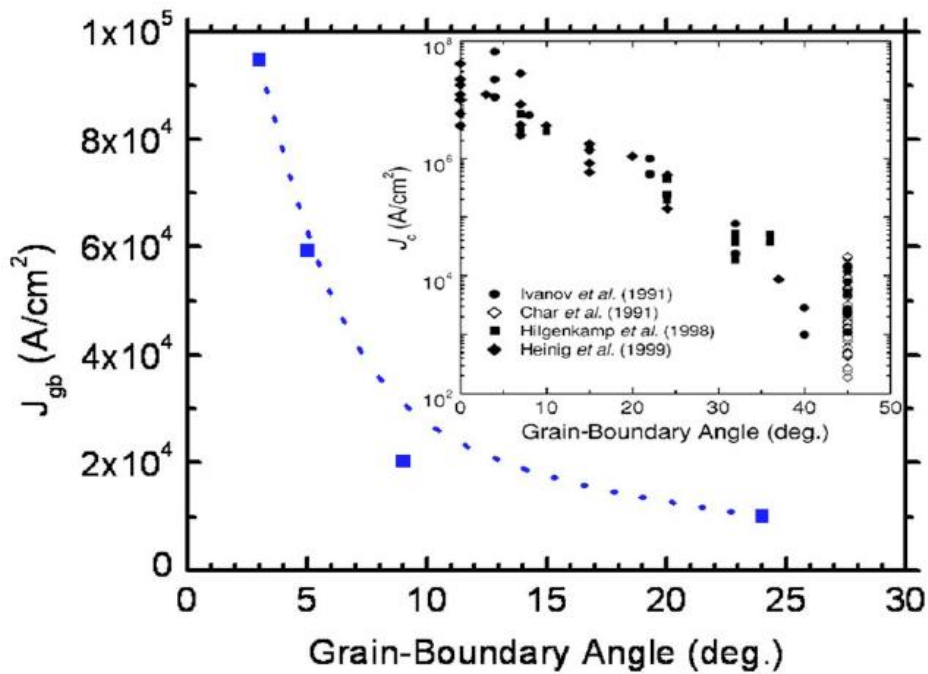

**Figure 27** Variation of intergranular $J_c$ with GB misorientation angle in Ba122:Co BGB junctions grown on [001]- tilt bicrystal substrates of MgO and LSAT. [78]

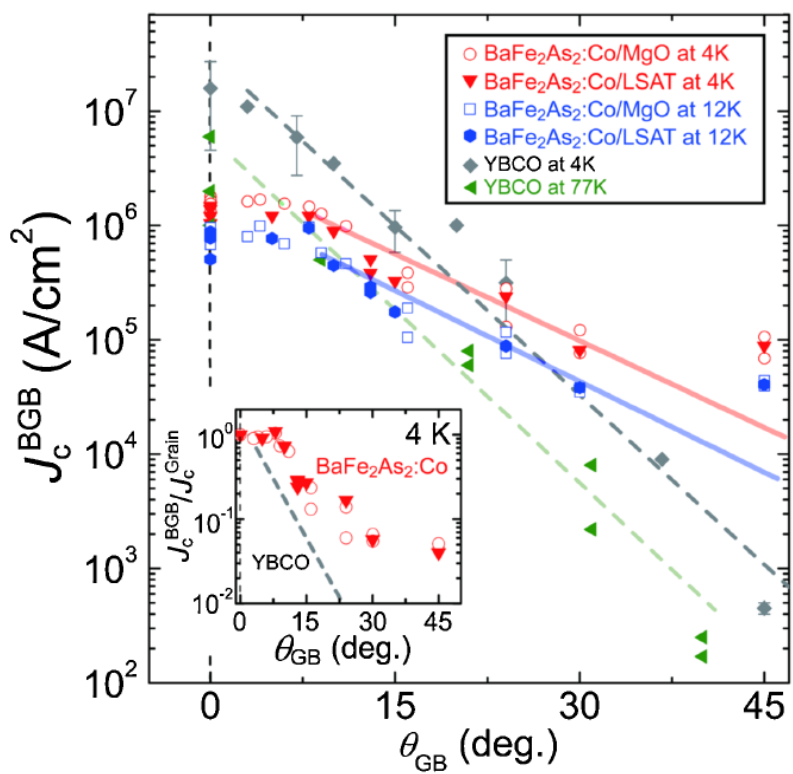

**Figure 28** The field dependence of transport $J_c$ at 4.2 K for pure and Sn added tapes [80]. The inset shows the $J_c(H)$ of the typical textured tape for parallel and perpendicular applied fields.

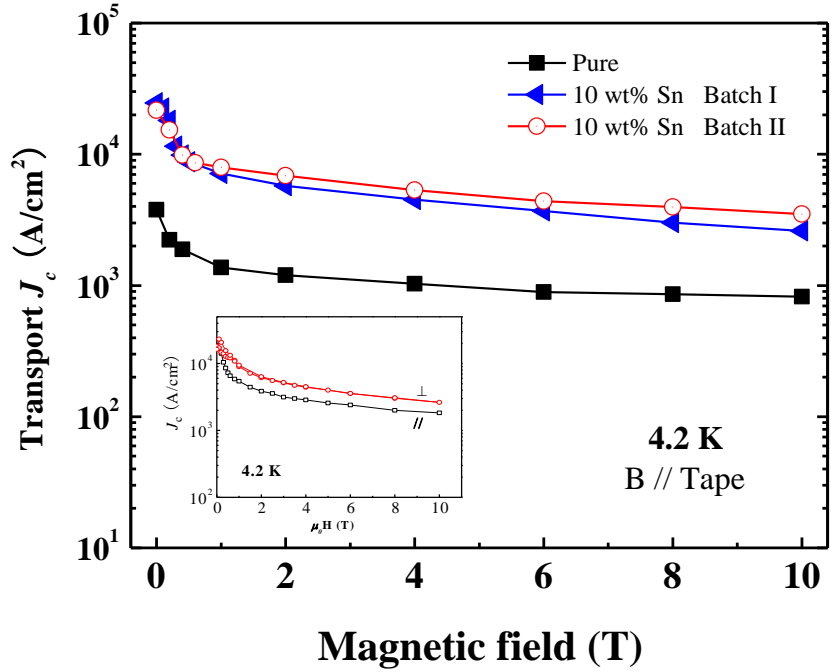

**Figure 29** The variation of $J_c(H)$ at 4.2 K for 122 pnictide tapes and wires fabricated by the *ex situ* PIT process. Data for the figure are taken from references given in the square brackets.

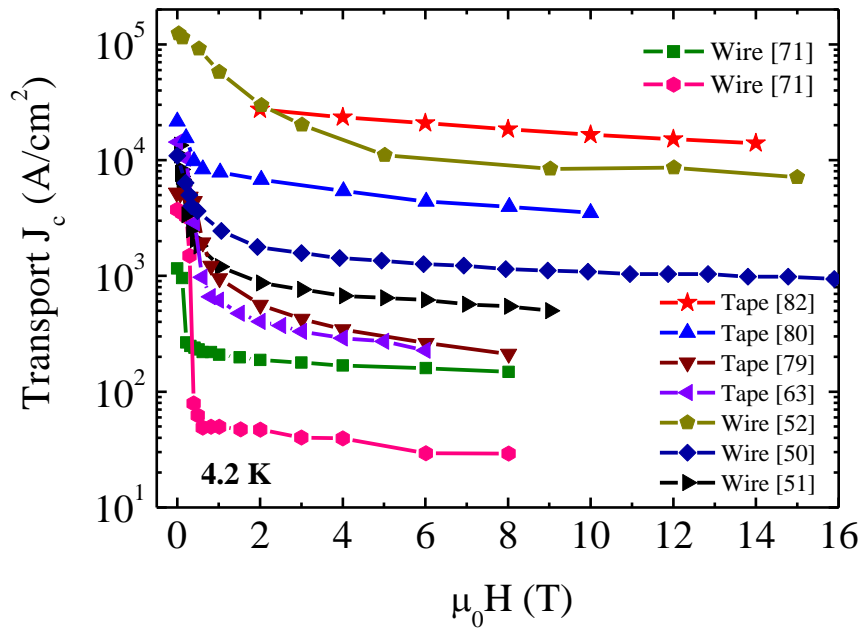

**Figure 30** Temperature dependence of resistivity for Sm-1111 polycrystals synthesized at different sintering temperatures from 850 to 1200°C [88].

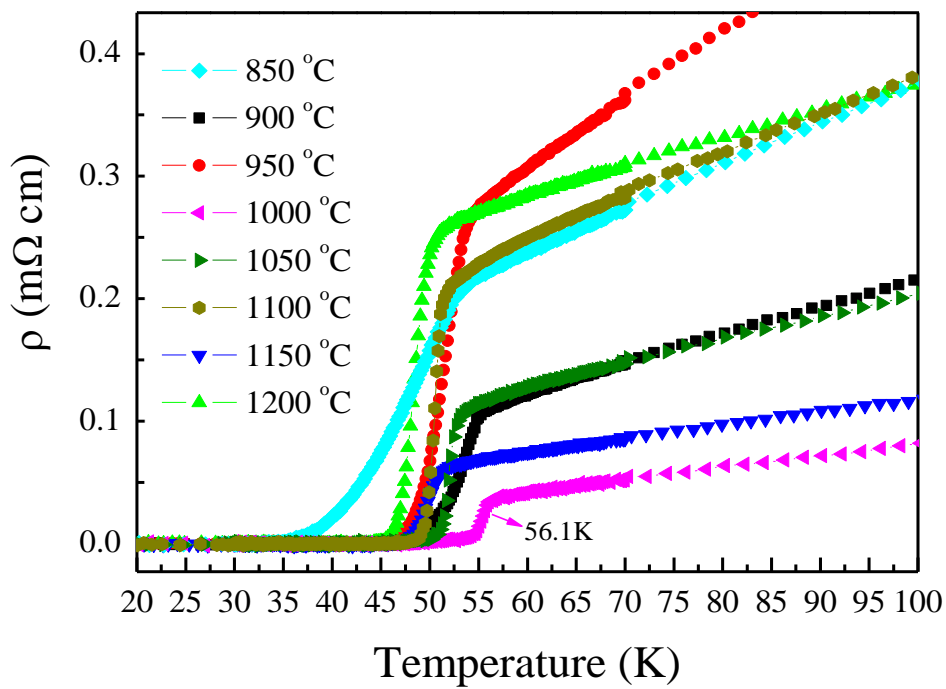

**Figure 31** Critical current density $J_c$ as a function of the applied magnetic field for Sm-1111 wires and tapes. Data for the figure are taken from references given in the square brackets.

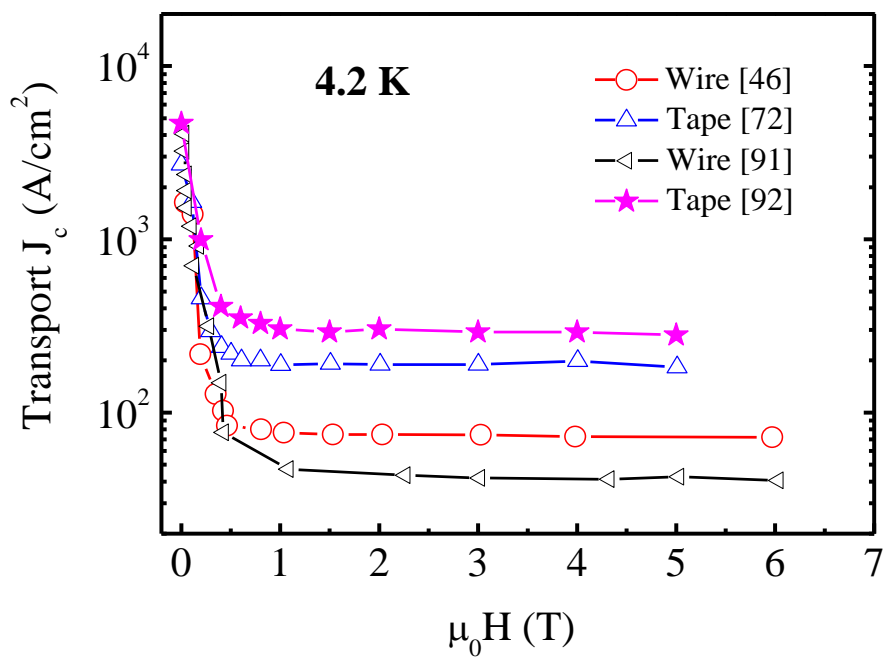

**Figure 32** SEM micrographs for the FeSe layers after peeling off the iron substrate. (c) shows photograph of the final FeSe wires and tapes. [95]

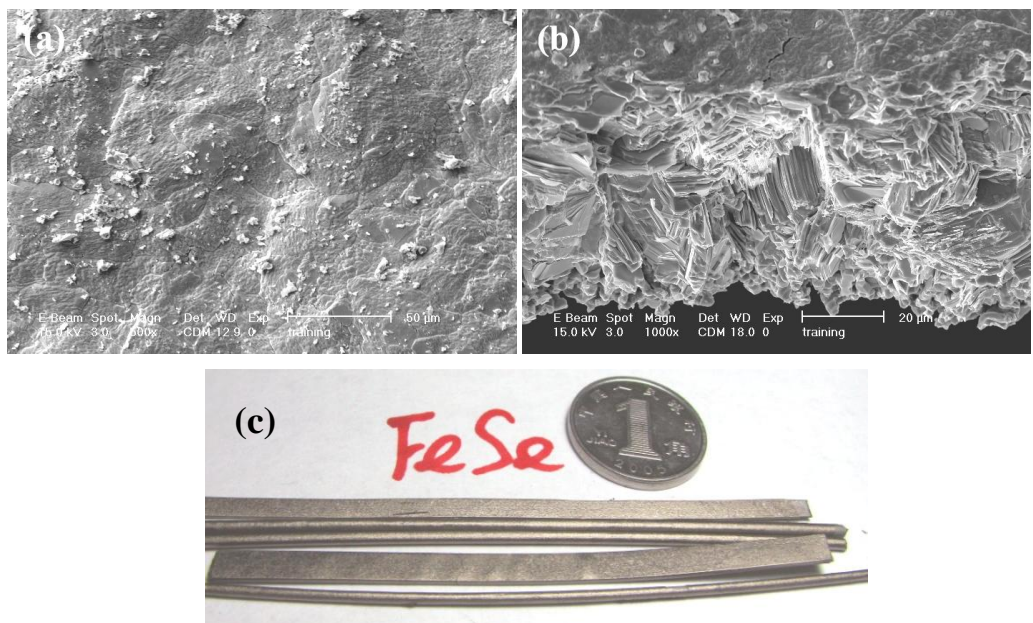

**Figure 33** Transport $J_c$ in applied fields of 11 type wires and tapes fabricated by different methods. Data for the figure are taken from references given in the square brackets.

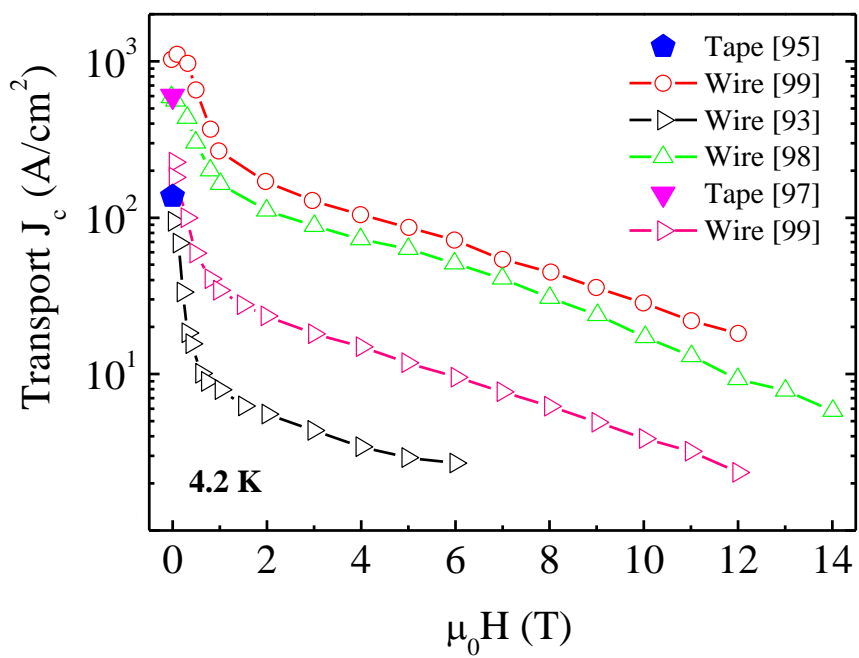

**Figure 34** Typical $J_c$-$H$ curve of the PIT processed Sr-122 wire, Co-doped Ba-122 film, Nb-Ti and Nb$_3$Sn superconducting wires and MgB$_2$ wire.

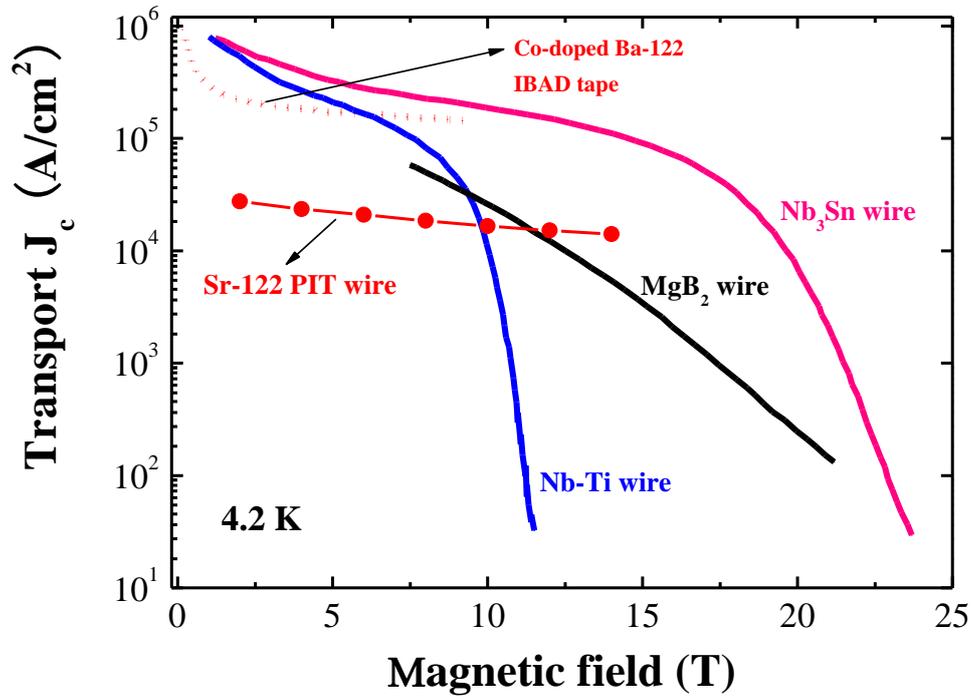